\documentclass[fleqn,usenatbib]{mnras}

\usepackage{newtxtext,newtxmath}
\usepackage{gensymb}
\usepackage{mathtools}
\usepackage{siunitx}

\usepackage[T1]{fontenc}

\DeclareRobustCommand{\VAN}[3]{#2}
\let\VANthebibliography\thebibliography
\def\thebibliography{\DeclareRobustCommand{\VAN}[3]{##3}\VANthebibliography}

\usepackage[flushleft]{threeparttable}
\usepackage{graphicx}	% Including figure files
\usepackage{amsmath}	% Advanced maths commands
\usepackage{pdflscape}% Extra maths symbols
\usepackage{xcolor}
\def\overbigdot#1{\overset{\hbox{\tiny$\bullet$}}{#1}}

\title[Synchrotron emission from V3890 Sgr]{Synchrotron emission from double-peaked radio light curves of the symbiotic recurrent nova V3890 Sagitarii}

\author[M. M. Nyamai et al.]{
Miriam~M.~Nyamai,$^{1,}$ $^{6}$\thanks{E-mail: nymmir001@myuct.ac.za (MMN)}
Justin~D.~Linford,$^{2}$
James~R.~Allison,$^{3,}$ $^{4}$
Laura~Chomiuk,$^{5}$
Patrick~A.~Woudt$^{6}$, 
\newauthor
Val\'erio~A.~R.~M.\ Ribeiro$^{7}$ and
Sumit~K.~Sarbadhicary$^{8}$
\\
$^{1}$South African Radio Observatory (SARAO), 2 Fir Street, Black River Park, Observatory, Cape Town, 7925, South Africa \\
$^{2}$National Radio Astronomy Observatory, P.O. Box O, Socorro, NM, 87801, USA\\
$^{3}$Sub-Deptartment of Astrophysics, Department of Physics, University of Oxford, Denys Wilkinson Building, Keble Rd., Oxford OX1 3RH, UK\\
$^{4}$First Light Fusion Ltd., Unit 9/10 Oxford Industrial Park, Mead Road, Yarnton, Kidlington OX5 1QU, UK\\
$^{5}$Center for Data Intensive and Time Domain Astronomy, Department of Physics and Astronomy, Michigan State University, East Lansing, MI 48824, USA\\
$^{6}$Department of Astronomy, University of Cape Town, Private Bag X3, Rondebosch 7701, South Africa\\
$^{7}$Instituto de Telecomunica\c{c}\~oes, Campus Universit\'ario de Santiago, 3810-193 Aveiro, Portugal\\
$^{8}$Center for Cosmology and AstroParticle Physics (CCAPP), The Ohio State University, 191 W. Woodruff Avenue, Columbus, OH 43210, USA\\
}

\date{Accepted XXX. Received YYY; in original form ZZZ}

\pubyear{2021}

\begin{document}
\label{firstpage}
\pagerange{\pageref{firstpage}--\pageref{lastpage}}
\maketitle

% Abstract of the paper
\begin{abstract}
We present radio observations of the symbiotic recurrent nova V3890 Sagitarii following the $2019$ August eruption obtained with the MeerKAT radio telescope at $1.28$ GHz and Karl G. Janksy Very Large Array (VLA) at $1.26~-35$ GHz. The radio light curves span from day $1$ to $540$ days after eruption and are dominated by synchrotron emission produced by the expanding nova ejecta interacting with the dense wind from an evolved companion in the binary system. The radio emission is detected early on (day $6$) and increases rapidly to a peak on day $15$. The radio luminosity increases due to a decrease in the opacity of the circumstellar material in front of the shocked material and fades as the density of the surrounding medium decreases and the velocity of the shock decelerates. Modelling the light curve provides an estimated mass-loss rate of ${\overbigdot M}_{\textrm {wind}} \approx 10^{-8} {\textrm {M}}_\odot~{\textrm {yr}}^{-1}$ from the red giant star and ejecta mass in the range of $M_{\textrm {ej}}=10^{-5}-10^{-6}~{\textrm {M}}_\odot$ from the surface of the white dwarf. 
V3890 Sgr likely hosts a massive white dwarf similar to other symbiotic recurrent novae, thus considered a candidate for supernovae type Ia (SNe Ia) progenitor. However, its radio flux densities compared to upper limits for SNe Ia have ruled it out as a progenitor for SN 2011fe. %V3890 Sgr hosts a massive white dwarf, shows presence of structured circumbinary material consistent with the surrounding environment present in some supernovae type Ia (SNe Ia) and therefore it is a potential SNe Ia progenitor.% In addition, we estimate a distance greater than $4.4$ kpc to the nova from kinematics of the Galaxy.
\end{abstract}

% Select between one and six entries from the list of approved keywords.
% Don't make up new ones.
\begin{keywords}
Radio continuum: transients;  novae, Cataclysmic variables; stars: individual: V3890 Sgr; acceleration of particles
\end{keywords}

%%%%%%%%%%%%%%%%%%%%%%%%%%%%%%%%%%%%%%%%%%%%%%%%%%

%%%%%%%%%%%%%%%%% BODY OF PAPER %%%%%%%%%%%%%%%%%%

\section{Introduction}

V3890 Sgr belongs to a class of cataclysmic variables known as symbiotic recurrent novae; ``symbiotic" because the accreting white dwarf has a giant companion \citep[e.g.,][]{schaefer2010}. %\citep{Munari2012}.
It is a recurrent nova because more than one thermonuclear eruption has been observed from this system; the most recent eruption occurred on $2019$ August $27.9$ UT as reported by A.\ Pereira \citep{strader2019, Kafka20}. Previous eruptions of the nova were observed in $1962$ June \citep{wenzel1990amplitude, miller1991v3890} and $1990$ April $27.7$ \citep[e.g.,][]{buckley1990v3890,Anupama1994}.

Initially, the orbital period of V3890 Sgr was estimated as $519.7$ days \citep{schaefer2009orbital}. However, recently using optical observations, \citet{2021MNRAS.504.2122M} estimated an orbital period  of $747.6$ days for V3890 Sgr. The nova is therefore, in the same class as other long period recurrent novae such as RS Oph, V745 Sco, T CrB and V2487 Oph \citep{anupama1999recurrent,schaefer2009orbital, schaefer2010}. Since some of the symbiotic recurrent novae have been extensively studied, their known parameters are listed in Table~\ref{tab:Table_novae} and compared with V3890 Sgr.

The optical light curves of these recurrent symbiotic novae evolve quickly following an eruption due to the low mass of material accreted onto the surface of the white dwarf since the last nova eruption, as these systems are known to host massive white dwarfs ($\gtrsim 1.2~\textrm{M}_\odot$; see Table~\ref{tab:Table_novae}). The ejecta outflows following an eruption are also fast, with velocities $\gtrsim 4000~\textrm {km~s}^{-1}$ measured from emission lines observed during different stages of the spectral evolution (see Table~\ref{tab:Table_novae}). The emission lines, however, become narrower with time during the early phase of the ejecta evolution, as the nova ejecta sweep up and are decelerated by the wind from the secondary star \citep[e.g.,][]{Riestra1992, Banerjee2014, Mondal2018}.  

%Based on the widths of the hydrogen Balmer emission lines, the nova envelope of RS Oph first undergoes a phase of free expansion for at least four days followed by a phase of deceleration lasting up to day $80$ after eruption \citep[see Figure $7$ of][]{Mondal2018}. A similar phenomenon is observed in V745 Sco, where the velocity implied from the FWHM of Pa$\beta$ emission lines decreases by a factor of about $3$ in $15$ days following the $2014$ eruption \citep{Banerjee2014}.

\begin{table*}
    \centering
    \caption{Estimated parameters of the symbiotic recurrent novae based on multiwavelength studies of individual systems. Listed parameters include masses of the binary components, the spectral classification of the companion star,  orbital period ($P_{\textrm{orbit}}$), years of recorded eruptions, the nova recurrence time ($t_{\textrm{rec}}$), velocity of the ejected material ($V_{\textrm{ej}}$), the time it takes for the nova to fade from optical maximum by $3$ magnitudes ($t_3$), and the distance to the nova ($d$).}
    \label{tab:Table_novae}
\begin{tabular}{lcccccccccr}
\hline
Name & $M_{\textrm{white dwarf}}$ & $M_{\textrm{giant}}$ & Spectral & $P_{\textrm{orbit}}$ & years of & $t_{\textrm{rec}}$ & $V_{\textrm{ej}}$ & $t_3$ & $d$  \\
& M$_{\odot}$  & M$_{\odot}$  &  type & (days)  & eruption & (yrs) & ($\textrm {km~s}^{-1}) $ &  (days) & (kpc)  \\
 \hline 
T CrB & 1.37 (1)  & 1.12 (1) & M4 III (2) & 228 (3) & 1866, 1946 & 80 & & 6 (8) &  0.81 (4) \\
RS Oph &1.2 -- & 0.7-- & M0--2 III (6) &  453.6 (5)  & 1898, 1907, 1933, 1945 & $\gtrsim 10$ & 4200 (7) & 14 (8) & 1.6 (9) \\
    & 1.4 (5)  & 0.8 (5)  &   &   & 1958, 1967, 1985, 2006, 2021 & & & & \\
V745 Sco &  &  & M4 III (10) & 510 (8) & 1937, 1963, 1989, 2014 & 25 & > 4000 (11) & 9 (8) & 7.8 (8) \\
V3890 Sgr & 1.35 (12) & 1.1 (12) & M5 III (10) & 747.6 (12) &1962, 1990, 2019 & 28 & $\gtrsim4200$ (13) & 14 (8) & 9 (12) \\
%V2487 Oph & & & & & 1900, 1998 & 98 &  & & &  \\
\hline 
\end{tabular}

\begin{tablenotes}
\item 
 References: (1) \citealp{Stanishev2004}; (2) \citealp{murset1999}; (3) \citealp{Kenyon1986}; (4) \citealp{Bailer2018}; (5) \citealp{Brandi2009}; (6) \citealp{anupama1999recurrent}; (7) \citealp{Mondal2018}; (8) \citealp{schaefer2010}; (9) \citealp{Hjellming1986}; (10) \citealp{harrison1993}; (11) \citealp{Banerjee2014}; (12)  \citealp{2021MNRAS.504.2122M}; (13) \citealp{strader2019};  (14) \citealp{Munari2019}

\end{tablenotes}
\end{table*}

Recurrent novae have short recurrence times of less than a century (see Table~\ref{tab:Table_novae}), which are attributed to high accretion rates, $\dot M \approx 10^{-8}~\textrm M_\odot~\textrm{ yr}^{-1}$
%, based on nova models 
\citep{Yaron2005}. The high rates rapidly supply enough material to power the subsequent thermonuclear runaway. Theoretically, most recurrent novae also consist of massive white dwarfs, and therefore require less mass to accumulate for hydrogen ignition \citep[e.g.,][]{Prialnik1995,Yaron2005, wolf2013}. In symbiotic systems, the high $\dot M$ is acquired through mass loss from the companion red giant, which is accreted either as a wind or through a disc via the inner Lagrangian point \citep{Luna2019}. The mass loss via the giant's wind also contributes to a dense circumstellar environment, which is impacted by the expanding nova envelope to give rise to shocks observed at high energies such as X-rays \citep[e.g.,][]{Bode1985, Sokoloski2006} and $\gamma-$rays \citep{Zheng2022}.% \citep{Abdo2010}. The first $\gamma-$ray detected nova was V407 Cyg, where the binary system is embedded in a wind of a Mira-type red giant companion star \citep{Abdo2010}.

A combination of high-mass white dwarfs and high mass accretion rates make the eruptions of these systems relatively gentle, and consequently not all of the accreted material is ejected during eruption \citep{Yaron2005}. The white dwarf may therefore grow in mass towards the Chandrasekhar limit. Indeed, the white dwarfs in recurrent novae have been shown to be massive \citep{Osborne2011, page2015}, and these systems have therefore been proposed as progenitors of supernovae type Ia  \citet{Maoz2014}. However, it is not clear whether the underlying white dwarfs in recurrent symbiotic novae are composed of CO or ONe. A CO white dwarf is required for a SN Ia; the fate of an ONe white dwarf that has grown in mass towards the Chandrasekhar limit is instead an accretion-induced collapse into a neutron star \citep{Gutierrez1996}.

Based on previous eruptions, the optical evolution of V3890 Sgr is fast, taking less than a day to rise to maximum magnitude ($V\approx~8$ mag) and $14$ days for the brightness to drop by $3$ mag; it is therefore classified as a fast nova \citep{payne1964galactic,schaefer2010}. 

The spectral evolution of V3890 Sgr at ultraviolet wavelengths shows the presence of both broad and narrow emission lines \citep{Riestra1992}. The broad lines originate from the expanding nova ejecta, while the narrow lines represent the speed of the red giant wind \citep[e.g.,][]{Munari2019c}. The FWHMs of the hydrogen Balmer lines decrease with time, from $2140~\textrm {km~s}^{-1}$ to $210~\textrm {km~s}^{-1}$ within a period of $13$ days following the eruption in 1990 \citep{Riestra1992}. Following the $2019$ eruption, the nova was observed with the Gemini observatory to obtain near-infrared spectra during the early days after the outburst \citep{Evans2022}. During this time, Helium emission lines showed both broad and narrow components. The broad lines emerged on day $3$ and narrowed through day $23$ \citep{Evans2022}. This is interpreted as evidence of the high velocity nova envelope being decelerated with time as it sweeps up the red giant wind.
%\subsection{The 2019 eruption of V3890 Sgr}
%\label{subsec:2019eruption}

\begin{figure}
  \centering
  \includegraphics[width=0.5\textwidth]{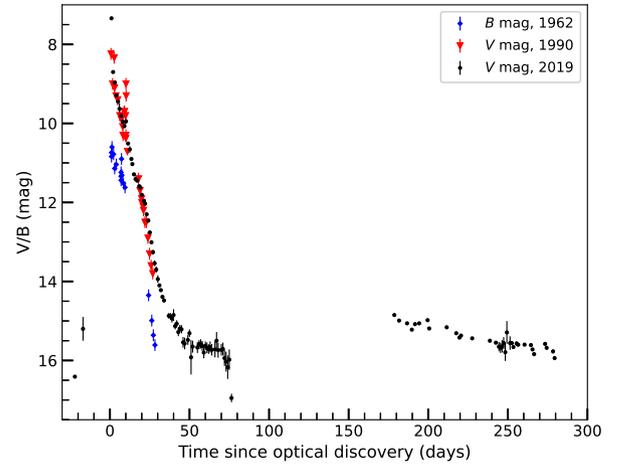}
  \caption{The $V$-band and $B$-band optical light curves of the $1962$, $1990$ and $2019$ eruptions of V3890 Sgr from AAVSO data \citep{Kafka20} and data presented by \citet{schaefer2010}. Here we set $t_0$ to time of the eruptions \citep{Sokolovsky2019}.}
  \label{fig:opticallightcurve}
\end{figure}

The rise, peak and decay of the optical light curve of V3890 Sgr following the $2019$ August eruption is well observed \citep{strader2019, Sokolovsky2019}. The optical evolution of the nova shown in Figure~\ref{fig:opticallightcurve} is similar to previous outbursts (see Figure~\ref{fig:opticallightcurve}).

The eruption has been observed at $\gamma-$ray, X-ray, infrared, UV, optical, and radio wavelengths \citep{buson2019, Orio2020, 2022Ness, 2022Kaminsky, Evans2022, nyamai2019radio, polisensky2019}. V3890 Sgr was detected in $\gamma$-rays and hard X-rays very soon ($\sim 2$ days) following the eruption \citep{buson2019, Sokolovsky2019}, consistent with expectations for a nova erupting in a dense environment. 

%The nova was detected with the VLA Low-band Ionosphere and Transient Experiment (VLITE) at $0.236~-~0.492$ GHz \citep{polisensky2019}.
Presented in this paper are radio observations of V3890 Sgr with MeerKAT at $1.28$ GHz and the VLA at radio frequencies between 1.26 GHz to $35$ GHz. The observations are used to study the transient phenomena of the system at radio frequencies. In \S \ref{sec:radio_observations}, radio observations and measurements including the radio light curve, radio spectral evolution and H{\sc~i} absorption analysis are presented. The emission from the nova, modelled as synchrotron radiation emanating from the interaction of the ejecta with the red giant wind, is analyzed in \S \ref{sec:synchrotron_discussion}. The conclusions are highlighted in \S \ref{sec:conclusions_V3890Sgr}.

\section{Radio observations}
\label{sec:radio_observations}

\subsection{MeerKAT and VLA observations of V3890 Sgr}
\label{sec:observations}

Monitoring of V3890 Sgr with the MeerKAT telescope started on ($t-t_0) \approx 2$ days, where $t_0$ is taken as $2019$ August $27.9$ (MJD 58722.9). V3890 Sgr is the first recurrent nova to be studied with MeerKAT. MeerKAT is a radio telescope located in South Africa and consists of $64$ dishes each with a diameter of $13.5$ m \citep{Jonas2016}. Combined, they form an array with a maximum baseline of $8$ km. The observations were taken using the MeerKAT L-band receiver, which has a total bandwidth of $856$ MHz split into $4096$ channels each with a width of $209$ kHz. %(\textcolor{red}{Is this still the case for MeerKAT?}).
The frequency range covered is $0.9$ to $1.67$ GHz centred at $1.284$ GHz. In each observation, the time on target was between $15$ and $30$ minutes (see Table~\ref{tab:Table_obs}). For the first $15$ days after optical discovery, V3890 Sgr was observed daily and then every two days afterwards until day $25$. Later, the observations were carried out once every week and finally the cadence was slowed down further to twice every month until the end of the observations on $2020$ May $26$. For all the epochs, the flux and bandpass calibrator J1939-6342 was observed for $\thicksim 5$ mins. The complex gain (secondary) calibrator J1911-2006 was observed for $2$ minutes per visit before and after observing the target.

%--Justin--
V3890 Sgr was observed with the VLA from $2019$ Aug $30$ to $2021$ Feb $13$ at observing frequencies between $1$ GHz and $37$ GHz. Observations were conducted using the L, C, Ku, and Ka band receivers. The total bandwidth was $1$ GHz for L band ($1-2$ GHz), $4$ GHz for C band ($4-8$ GHz), $6$ GHz for Ku band ($12-18$ GHz), and $8$ GHz for Ka band ($28.5-32.5$ and $34.0-38.0$ GHz). For all epochs, the absolute flux density and bandpass calibrator 3C286 was observed for $2$ minutes per band. The complex gain calibrator J1820-2528 was used for all observations at C, Ku, and Ka bands. For the L band observations, the complex gain calibrator was J1820-2528 during A and B configurations, and J1833-2103 during C and D configurations. VLA observations were carried out under programs VLA/19B-313 and VLA/20B-302. The total observing time on target aross all frequency bands varied between $17$ mins and $78$ mins (see Table~\ref{tab:Table_obs}).

\begin{table*}
    \centering
    \caption{A summary of MeerKAT and VLA observations of V3890 Sgr.}
    \label{tab:Table_obs}
\begin{tabular}{lccccr}

\hline 
Observation & $t$ & $t-t_0$ & Observation frequency & Telescope & Observation time\\
Date & (MJD) & (Days) &  (GHz) &  & (mins)\\ 
\hline 

2019 Aug 29 & 58724.8 & 1.9 &1.28 & MeerKAT & 30.0 \\
2019 Aug 30 & 58725.7 & 2.8 & 1.28 & MeerKAT & 30.0\\
2019 Aug 30 & 58726.0 & 3.1 &  5.0, 7.0 & VLA (A config.) & 18.0 \\
2019 Aug 31 & 58726.8 & 3.9 & 1.28 & MeerKAT & 30.0 \\
2019 Sep 01 & 58727.8  & 4.9 & 1.28 & MeerKAT & 30.0 \\
2019 Sep 02 & 58728.8 & 5.9 & 1.28 & MeerKAT & 30.0 \\
2019 Sep 03 & 58729.8  & 6.9 & 1.28 & MeerKAT & 30.0 \\
2019 Sep 04 & 58730.0 &  7.2 & 1.26, 1.78, 5.0, 7.0 & VLA (A config.) & 18.0 \\ 
2019 Sep 04  & 58730.9 & 8.0 &  1.28 & MeerKAT & 30.0 \\ 
2019 Sep 05 & 58731.0 & 8.1  & 1.26, 1.78, 5.0, 7.0 & VLA (A config.) & 18.0 \\
2019 Sep 05 & 58731.9 & 8.5 & 1.28 & MeerKAT & 30.0 \\
2019 Sep 06 & 58732.9 & 9.5 & 1.28 & MeerKAT & 30.0 \\
2019 Sep 07 & 58733.6 & 10.4 & 1.28 & MeerKAT & 30.0 \\
2019 Sep 08 & 58734.1 & 11.2 &  1.26, 1.78, 5.0, 7.0 & VLA (A config.) & 17.0 \\
2019 Sep 08 & 58734.6 & 11.7  & 1.28 & MeerKAT & 30.0 \\
2019 Sep 09 & 58735.8 & 12.9 & 1.28 & MeerKAT & 15.0 \\
2019 Sep 11 & 58737.0 & 14.1 & 1.26, 1.78, 5.0, 7.0 & VLA (A config.) & 17.0 \\
2019 Sep 11 & 58737.8 & 14.9 & 1.28 & MeerKAT & 15.0 \\
2019 Sep 13 & 58739.0 & 16.1 & 1.26, 1.78, 5.0, 7.0 & VLA (A config.) & 17.0 \\
2019 Sep 13 & 58739.9  & 17.0 & 1.28 & MeerKAT & 15.0 \\
2019 Sep 15 & 58741.8 & 18.9 & 1.28 & MeerKAT & 15.0 \\
2019 Sep 17 & 58743.9 & 21.0&  1.28 & MeerKAT & 15.0 \\
2019 Sep 18 & 58744.1 & 21.2 & 1.26, 1.78, 5.0, 7.0 & VLA (A config.) & 17.0 \\
2019 Sep 19 & 58745.8 & 22.9&  1.28 & MeerKAT & 15.0 \\
2019 Sep 21 & 58747.6 & 24.7& 1.28 & MeerKAT & 15.0  \\
2019 Sep 22 & 58749.0 & 26.1 & 1.26, 1.78, 5.0, 7.0 & VLA (A config.) & 17.0 \\
2019 Sep 24 & 58751.0 & 28.1 & 1.26, 1.78, 5.0, 7.0 & VLA (A config.) & 17.0\\
2019 Sep 26 & 58752.7 & 29.8& 1.28 & MeerKAT & 15.0 \\
2019 Sep 29 & 58755.7 & 32.8& 1.28 & MeerKAT & 15.0 \\
2019 Sep 30 & 58756.9 & 34.0 &  1.26, 1.78, 5.0, 7.0 & VLA (A config.) & 17.0 \\
2019 Oct 04 & 58761.0 & 38.1 & 1.26, 1.78, 5.0, 7.0 & VLA (A config.) & 17.0 \\
2019 Oct 06 & 58762.7 & 39.8 & 1.28 & MeerKAT & 15.0 \\
2019 Oct 08 & 58765.0 & 42.1 & 1.26, 1.78, 5.0, 7.0 & VLA (A config.) & 17.0 \\
2019 Oct 11 & 58768.0 & 45.0 & 1.26, 1.78, 5.0, 7.0 & VLA (A config.) & 17.0 \\
2019 Oct 17 & 58774.0 & 51.1 & 1.26, 1.78, 5.0, 7.0 & VLA (A config.) & 17.0 \\
2019 Oct 26 & 58782.6 & 59.7 & 1.28 & MeerKAT & 20.0  \\
2019 Nov 02 & 58790.0 & 67.0 & 1.26, 1.78, 5.0, 7.0, 13.5, 16.5, 29.5, 35.0 & VLA (A config.) & 40.6 \\
2019 Nov 18 & 58805.7 & 82.8 & 1.28 & MeerKAT & 20.0 \\
2019 Nov 18 & 58806.8 & 83.9 & 1.26, 1.78, 5.0, 7.0 & VLA (D config.) & 17.0\\
2019 Nov 23 & 58810.9 & 88.0 & 13.5, 16.5, 29.5, 35.0 & VLA (D config.) & 24.3 \\
2019 Nov 30 & 58817.6 & 94.7 & 1.28 & MeerKAT & 20.0 \\
2019 Dec 10 & 58827.8 & 104.9 & 1.26, 1.78, 5.0, 7.0, 13.5, 16.5, 29.5, 35.0 & VLA (D config.) & 37.9 \\
2019 Dec 13 & 58830.6 & 107.7 & 1.28 & MeerKAT & 20.0 \\
2020 Jan 04 & 58852.8 & 129.8 & 5.0, 7.0, 13.5, 16.5, 29.5, 35.0 & VLA (D config.) & 37.9\\
2020 Feb 26 & 58905.5 & 182.6 & 1.26, 1.74, 13.5, 16.5, 29.5, 35.0  & VLA (D config.) & 37.9\\
2020 Mar 24 & 58932.1 & 209.2 & 1.28 & MeerKAT & 15.0 \\
2020 April 08 & 58947.4 & 224.5 & 1.26, 1.74, 13.5, 16.5, 29.5, 35.0 & VLA (C config.) & 37.9\\
2020 Apr 14 & 58953.0 & 230.1 & 1.28 & MeerKAT & 15.0 \\
2020 May 18 & 58988.0 & 265.1 & 1.28 & MeerKAT & 15.0 \\
2020 May 21 & 58990.3 & 267.4 & 1.26, 1.78, 13.5, 16.5, 29.5, 35.0 & VLA (C config.) & 37.9 \\
2020 May 26 & 58995.2 & 272.3 & 1.28 & MeerKAT & 15.0 \\
2020 Jul 20 & 59050.2 & 327.3 & 1.26, 1.78, 13.5, 16.5, 29.5, 35.0 & VLA (B config.) & 38.6 \\
2020 Oct 06 & 59128.0 & 405.1 & 1.26, 1.78, 13.5, 16.5, 29.5, 35.0 & VLA (B config.) & 44.5 \\
2021 Feb 13 & 59258.6 & 535.7 & 1.26, 1.78, 5.0, 7.0 & VLA (A config.) & 78.4 \\

\hline 
\end{tabular}

\begin{tablenotes}
\item Here, $t_0$ is taken as 2019 August 27.9 UT (MJD = 58722.9). \end{tablenotes}
\end{table*}

\subsection{Data reduction}
\label{sec:data}

Data reduction for MeerKAT was undertaken using \texttt{CASA} \citep{McMullin2007}. To remove radio frequency interference, the data were flagged using the \texttt{AOflagger} algorithm \citep{offringa2012}. In order to find the bandpass corrections, first the phase-only and antenna-based delay corrections were determined on the primary calibrator. The primary calibrator bandpass corrections were then applied. This was followed by solving for complex gains for both the primary and secondary calibrators. The absolute flux density of the secondary calibrator was estimated by scaling the corrections from the primary to the secondary calibrator. Calibrations and the absolute flux density scale were consequently transferred to V3890 Sgr (the target). 

Imaging of MeerKAT data was performed using \texttt{WSCLEAN} \citep{offringa2014} using a Briggs weighting with a robust value of $-0.7$. These steps are summarised in the oxkat pipeline \footnote{https://www.
github.com/IanHeywood/oxkat and through the Astrophysics Source Code Library record ascl:2009.003 \citep{Heywood2020}.}. The flux densities of V3890~Sgr were estimated with the \texttt{CASA IMFIT} task by fitting a Gaussian to the image of the target. Since the nova was very bright, exhibiting a high signal to noise ratio, the width of the Gaussian was allowed to vary when obtaining the flux density measurements. For non-detections, the upper limit was calculated as the pixel value at the location of the nova added to $3~\times$ rms value of a region in the image away from the target location. The observations and results are presented in Table~\ref{tab:Table_results} and plotted in Figure~\ref{fig:radio_light_curve}. The quoted errors on the flux densities include Gaussian fit errors ($\sigma_{\textrm{Fit}}$) and $10\%$ calibration errors ($\sigma_{\textrm{cal}} = 0.1S_\nu$; e.g., \citealt{Hewitt2020}) such that
\begin{equation}
\sigma_{S_\nu} = \sqrt{\sigma_{\textrm{Fit}}^2~+~(0.1\times~S_\nu)^2}.
\label{eq:uncertanity}
\end{equation}

%--Justin --
The VLA data were calibrated using the VLA \texttt{CASA} calibration pipeline versions $5.4.2$ and $5.6.2$.  Additional flagging was done in \texttt{AIPS} \citep{2003ASSL..285..109G}, and final imaging was done using \texttt{Difmap} \citep{1997ASPC..125...77S}. The VLA data for each band were split into two subbands for imaging to increase spectral coverage. When possible, phase self-calibration was performed in \texttt{Difmap} as part of the imaging process. All images where the nova was detected were then loaded into \texttt{AIPS} and flux density measurements were made using the \texttt{JMFIT} task. As with the MeerKAT data, we include 10\% calibration uncertainty for the VLA flux density values, added in quadrature with the statistical uncertainty from \texttt{JMFIT}.  In the cases of non-detections, we recorded the flux density value at the location of the nova and obtained the image rms in a region away from the nova.  The upper limit values presented for the VLA are the flux density value at the nova location plus $3~\times$ the off-source image rms.

\subsection{Radio light curve}
\label{sec:lightcurveV3890Sgr}

\begin{figure*}
  \centering
  \includegraphics[width=0.98\textwidth]{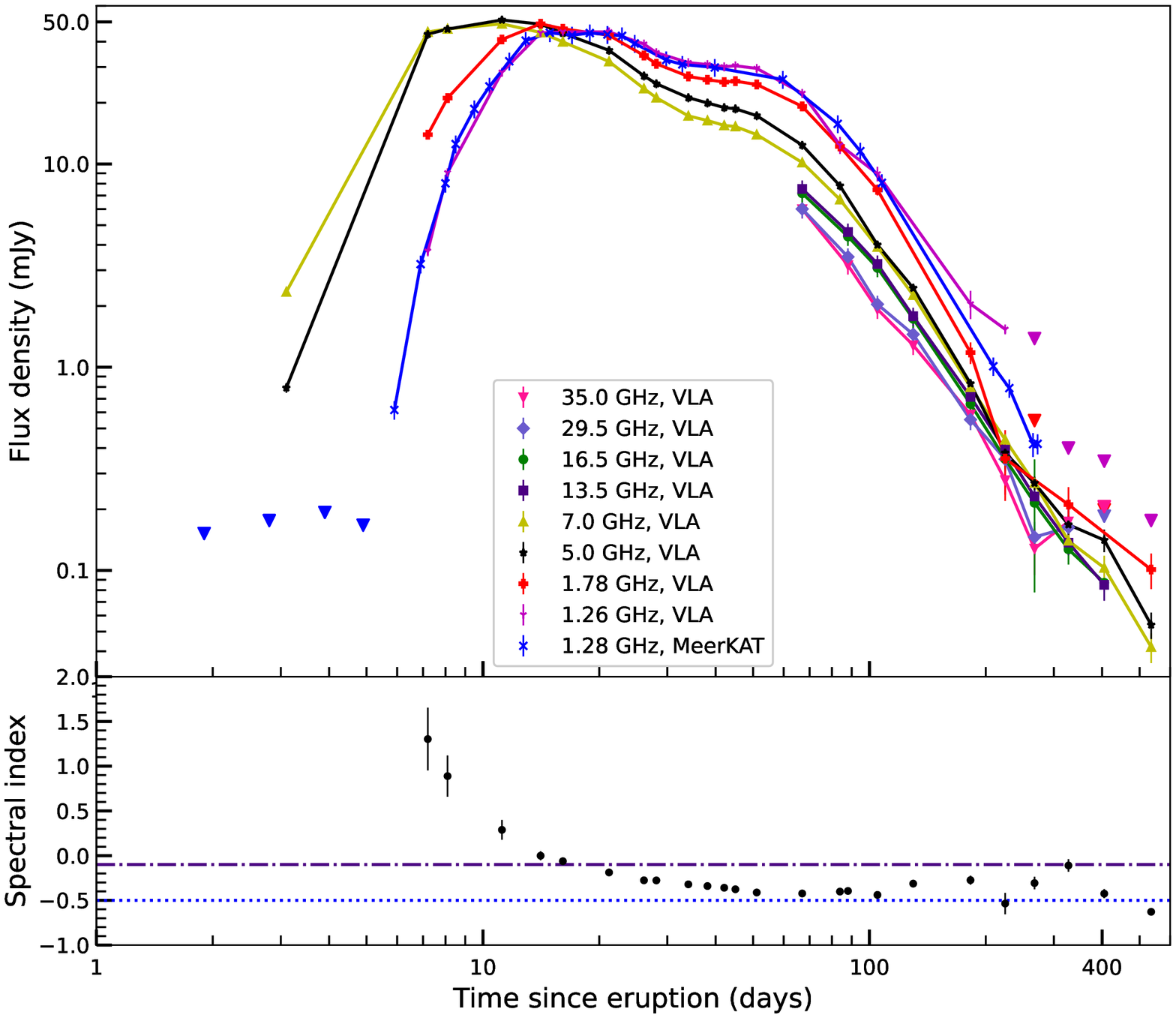}
  \caption{Top: Observed flux densities of V3890~Sgr spanning from day $2$ to day $536$ after the nova eruption. The upper limits in flux density are included as triangles. We take MJD = $58722.9$ as the date of the eruption ($t_0$). Bottom: spectral indices obtained by fitting a single power-law to multi-band data ($1.26-35$) GHz. The dash-dotted line in the lower panel represents $\alpha = -0.1$, the theoretically expected index of optically thin free-free emission. The dotted line represents $\alpha = -0.5$, the expected index of optically thin synchrotron emission.}
  \label{fig:radio_light_curve}
\end{figure*}

%V3890 Sgr is the first recurrent nova to be studied with MeerKAT.
The $1$ to $37$ GHz light curves are plotted in Figure~\ref{fig:radio_light_curve}. Initially, during day $2$ to day $5$ after eruption, the nova was not detected, with $3\sigma$ upper limits of $<0.2$ mJy. V3890 Sgr then shows a rapid increase in flux density, which peaked first on day $19$ and again on day $60$, forming a double peaked radio light curve. The flux density varies with frequency, such that the light curve peaks first at higher frequencies (> $5$ GHz). During this time, the flux is still rising at lower frequencies (< $2$ GHz), which peak later. The amplitude of the peaks vary with frequency, such that during the first peak, (day $19$), the brightness remains the same at all observed frequencies. However, during the second peak, (day $60$), the nova is brighter at lower frequencies. After the secondary peak, the nova faded to flux densities below $0.1$ mJy.

\begin{landscape}
\begin{table}
    \centering
    \caption{A summary of the radio flux densities of V3890 Sgr from MeerKAT and VLA observations.}
    \label{tab:Table_results}
\begin{tabular}{lcccccccccr}

\hline 
$t-t_0$ & $S\textunderscore1.28$ (mJy) & $S\textunderscore1.26$ (mJy) & $S\textunderscore1.78$ (mJy) & $S\textunderscore5.0$ (mJy) & $S\textunderscore7.0$ (mJy) & $S\textunderscore13.5$ (mJy) & $S\textunderscore16.5$ (mJy) & $S\textunderscore29.5$ (mJy) & $S\textunderscore35.0$ (mJy) & Spectral\\
(Days) & &  &  & & & & & & & index ($\alpha$)\\ 
\hline

1.9 & $<$ 0.15  &  &  & & & & & & & \\ 
2.8 & $<$ 0.18  &  &  & & & & & & & \\ 
3.1 & & &  & 0.79 $\pm$ 0.04 & 2.35 $\pm$ 0.12& & & & & \\ 
3.9 &  $<$ 0.19 & &  & & & & & & & \\ 
4.9 &  $<$ 0.18  & &  & & & & & & & \\
5.9 &  0.62 $\pm$ 0.06 & &  & & & & & & & \\
6.9 &  3.22 $\pm$ 0.32 & &  & & & & & & & \\
7.2 & & 3.73 $\pm$ 0.21 & 13.94 $\pm$ 0.70 & 43.48 $\pm$ 2.17 & 44.62 $\pm$ 2.23 & & & & & 1.30 $\pm$ 0.35 \\
8.0 &  8.03 $\pm$ 0.80 & &  & & & & & & & \\
8.1 &  & 9.06 $\pm$ 0.46  & 21.10 $\pm$ 1.06 & 46.09 $\pm$ 2.31 & 46.17 $\pm$ 2.31 & & & & & 0.90 $\pm$ 0.22\\
8.5 & 12.53 $\pm$ 1.25 & &  & & & & & & & \\
9.5 & 18.65 $\pm$ 1.87 & &  & & & & & & & \\
10.4 & 24.08 $\pm$ 2.41  & &  & & & & & & & \\
11.2 & & 28.30 $\pm$ 1.42 & 40.97 $\pm$ 2.05 & 51.11 $\pm$ 2.56 & 48.93 $\pm$ 2.45 & & & & & 0.29 $\pm$ 0.11 \\
11.7 & 32.03 $\pm$ 3.21 & &  & & & & & & & \\
12.9 & 40.44 $\pm$ 4.05 & &  & & & & & & & \\
14.1 & & 44.15 $\pm$ 2.21 & 48.95 $\pm$ 2.45 & 48.74 $\pm$ 2.44  & 44.35 $\pm$ 2.22 & & & & & 0.00 $\pm$ 0.05 \\
14.9 & 44.20 $\pm$ 4.42 & &  & & & & & & & \\
16.1 & & 44.67 $\pm$ 2.24 & 46.26 $\pm$ 2.31 & 44.01 $\pm$ 2.20 & 39.89 $\pm$ 2.00 & & & & & $-$0.06 $\pm$ 0.03 \\
17.0 & 43.0 $\pm$ 4.30 & &  & & & & & & & \\
18.9 & 44.07 $\pm$ 4.41 & & & & & & & & & \\
21.0 & 43.35 $\pm$ 4.34 & & & & & & & & & \\
21.2 &  & 44.66 $\pm$ 2.24 & 42.95 $\pm$ 2.15 & 36.23 $\pm$ 1.81 & 31.92 $\pm$ 1.60 & & & & & $-$0.19 $\pm$ 0.02 \\
21.9 & 42.68 $\pm$ 4.27 & &  & & & & & & & \\
24.7 & 39.18 $\pm$ 3.92 & &  & & & & & & & \\
26.1 &  & 38.89 $\pm$ 1.95 & 34.18 $\pm$ 1.71 & 27.12 $\pm$ 1.36 & 23.49 $\pm$ 1.18 & & & & & $-$0.28 $\pm$ 0.02 \\
28.1 &  & 35.30 $\pm$ 1.77 & 31.10 $\pm$ 1.56 & 24.83 $\pm$ 1.24 & 21.18 $\pm$ 1.06 & & & & & $-$0.28 $\pm$ 0.03 \\
29.8 & 32.49 $\pm$ 3.25 & &  & & & & & & & \\
32.8 & 30.88 $\pm$ 3.10 & &  & & & & & & & \\
34.0 & & 31.53 $\pm$ 1.58 & 26.97 $\pm$ 1.35  & 21.19 $\pm$ 1.06 & 17.24 $\pm$ 0.86 & & & & & $-$0.32 $\pm$ 0.04 \\
38.1 & & 30.85 $\pm$ 1.54  & 25.99 $\pm$ 1.30 & 19.96 $\pm$ 1.00 & 16.36 $\pm$ 0.82 & & & & & $-$0.34 $\pm$ 0.04 \\
39.8 & 29.84 $\pm$ 2.99 & &  & & & & & & & \\
42.1 & & 30.09 $\pm$ 1.51 & 25.30 $\pm$ 1.27 & 18.95 $\pm$ 0.95 & 15.48 $\pm$ 0.77 & & & & & $-$0.36 $\pm$ 0.03 \\
45.0 & & 30.41 $\pm$ 1.52  & 25.55 $\pm$ 1.28  & 18.68 $\pm$ 0.93  & 15.28 $\pm$ 0.76 & & & & & $-$0.38 $\pm$ 0.03 \\
51.1 &  & 29.53 $\pm$ 1.48  & 24.62 $\pm$ 1.23 & 17.29 $\pm$ 0.86 & 13.94 $\pm$ 0.70 & & & & & $-$0.41 $\pm$ 0.03 \\
59.7 & 26.04 $\pm$ 2.61  & &  & & & & & & & \\
67.0 &   &  22.26 $\pm$ 1.12 & 19.21 $\pm$ 0.96 & 12.38 $\pm$ 0.62 & 10.17 $\pm$ 0.51 & 7.54 $\pm$ 0.75 & 7.17 $\pm$ 0.72  & 6.01 $\pm$ 0.60  & 5.99 $\pm$ 0.60  & $-$0.42 $\pm$ 0.01 \\
82.8 & 15.76 $\pm$ 1.58  & &  & & & & & & & \\
83.9 &   & 12.39 $\pm$ 1.19  & 12.19 $\pm$ 0.69  & 7.81 $\pm$ 0.39 & 6.68 $\pm$ 0.34 & & & & & $-$0.40 $\pm$ 0.04 \\
88.0 &  &  &  & & & 4.62 $\pm$ 0.46 & 4.40 $\pm$ 0.44 & 3.49 $\pm$ 0.35 & 3.18 $\pm$ 0.32 & $-$0.39 $\pm$ 0.03\\
94.7 & 11.56 $\pm$ 1.16  & &  & & & & & & & \\
104.9 &  & 8.94 $\pm$ 0.74  & 7.46 $\pm$ 0.45 & 4.0 $\pm$ 0.20 & 3.9 $\pm$ 0.20 & 3.22 $\pm$ 0.32 & 3.08 $\pm$ 0.31 & 2.04 $\pm$ 0.21 & 1.93 $\pm$ 0.20 & $-$0.44 $\pm$ 0.03 \\
107.7 & 8.08 $\pm$ 0.81   & &  & & & & & & & \\

\hline 

\end{tabular}

\begin{tablenotes}
\item Here, $t_0$ is taken as 2019 August 27.9 UT (MJD = 58722.9). \end{tablenotes}
\end{table}
\end{landscape}

\begin{landscape}
\begin{table}
    \centering
    \contcaption{A summary of MeerKAT and VLA observations of V3890 Sgr.}
    %\label{tab:Table_results}
\begin{tabular}{lcccccccccr}

\hline 
$t-t_0$ & $S\textunderscore1.28$ (mJy) & $S\textunderscore1.26$ (mJy) & $S\textunderscore1.78$ (mJy) & $S\textunderscore5.0$ (mJy) & $S\textunderscore7.0$ (mJy) & $S\textunderscore13.5$ (mJy) & $S\textunderscore16.5$ (mJy) & $S\textunderscore29.5$ (mJy) & $S\textunderscore35.0$ (mJy) & Spectral\\
(Days) & &  &  & & & & & & & index ($\alpha$)\\ 
\hline
129.8 &  & &  & 2.45 $\pm$ 0.13  & 2.26 $\pm$ 0.12 & 1.78 $\pm$ 0.18 & 1.73 $\pm$ 0.17 & 1.45 $\pm$ 0.15  & 1.28 $\pm$ 0.13 & $-$0.32 $\pm$ 0.01\\
182.6 & & 2.05 $\pm$ 0.32 & 1.18 $\pm$ 0.14  & 0.83 $\pm$ 0.04 & 0.80 $\pm$ 0.04 & 0.72 $\pm$ 0.07 & 0.66 $\pm$ 0.07 & 0.55 $\pm$ 0.06& 0.59 $\pm$ 0.07 & $-$0.28 $\pm$ 0.05 \\
209.2  & 1.01 $\pm$ 0.11 & &  & & & & & & & \\
224.5 &  & 1.54 $\pm$ 0.08 & 0.36 $\pm$ 0.14 & 0.38 $\pm$ 0.02 & 0.44 $\pm$ 0.03 & 0.39 $\pm$ 0.04 & 0.35 $\pm$ 0.04 &  0.35 $\pm$ 0.05& 0.28 $\pm$ 0.04 & $-$0.10 $\pm$ 0.06\\
230.1 &  0.79 $\pm$ 0.08 & &  & & & & & & & \\
265.1 & 0.42 $\pm$ 0.06  & &  & & & & & & & \\
267.4 & & $<$1.38 & $<$ 0.55 & 0.27 $\pm$ 0.02& 0.27$\pm$ 0.02 & 0.23 $\pm$ 0.03 & 0.22 $\pm$ 0.14 & 0.15 $\pm$ 0.03  & 0.13 $\pm$ 0.03  & $-$0.32 $\pm$ 0.07 \\
272.3  & 0.42 $\pm$ 0.05 & &  & & & & & & & \\
327.3 & & $<$ 0.40 & 0.21 $\pm$ 0.05 & 0.17 $\pm$ 0.02 &0.14 $\pm$ 0.02 & 0.14 $\pm$ 0.02 & 0.13 $\pm$ 0.02 & 0.16 $\pm$ 0.04 & 0.17 $\pm$ 0.05 & $-$0.11 $\pm$ 0.07\\
405.1 & & $<$ 0.35 & $<$ 0.55 & 0.14 $\pm$ 0.02 & 0.10 $\pm$ 0.02 & 0.09 $\pm$ 0.01 & 0.09 $\pm$ 0.01 & $<$ 0.19 & $<$ 0.21 & $-$0.41 $\pm$ 0.11 \\
535.7  & & $<$0.18 & 0.10 $\pm$ 0.02  & 0.05 $\pm$ 0.01 & 0.04 $\pm$ 0.01 & & & & & $-$0.63 $\pm$ 0.01 \\
\hline 

\end{tabular}

\begin{tablenotes}
\item Here, $t_0$ is taken as 2019 August 27.9 UT (MJD = 58722.9). \end{tablenotes}
\end{table}
\end{landscape}

\subsection{Radio spectral evolution}
\label{sec:spectra_V3890Sgr}

\begin{figure*}
  \centering
  \includegraphics[width=0.98\textwidth]{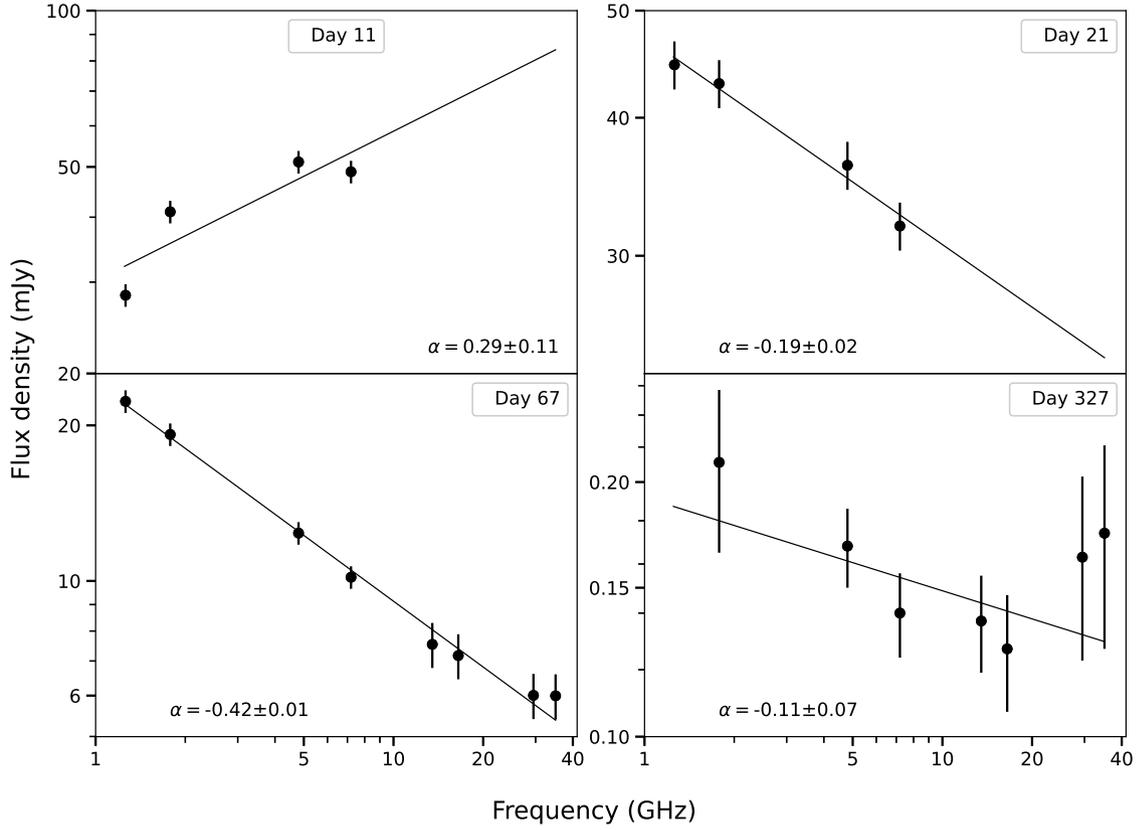}
  \caption{Spectral evolution of V3890 Sgr after the $2019$ eruption. The time since eruption and $\alpha$ is given for each fit represented by a solid line.}
  \label{fig:spectra_V3890_Sgr}
\end{figure*}

To determine the spectral evolution of V3890 Sgr, the data are fit assuming a simple power law using the method of least-squares. Measurements of the spectral index ($\alpha$ where $S_\nu~\propto~\nu^{\alpha}$) are presented in  Figure~\ref{fig:radio_light_curve} and Figure~\ref{fig:spectra_V3890_Sgr}. During the early times (e.g., on day $11.2$), the flux density rises towards higher frequency, giving $\alpha = 0.29$, an indicator of optically thick emission. Around the radio light curve maximum, e.g., on day $21$ the spectrum switches to rising towards lower frequency, and is well described by a power law with slope $\alpha = -0.2$.

%Similar values of the spectral index, $\alpha= -0.3$ are estimated between $1.26-7.0$ GHz using the Karl G.Jansky Very Large Array and VLA Low-band Ionosphere and Transient Experiment (VLITE) \citep{polisensky2019}.

The spectrum continues to steepen ($\alpha$ becomes more negative,) with time, as shown on day $67$ where $\alpha = -0.4$. On day $327$, the spectrum is flat $\alpha= -0.1$) indicative of optically thin free-free emission, however on day $534$, the spectrum rises steeply again towards lower frequencies and can be fit with a single power-law such that $\alpha = -0.6$, an indication of optically thin synchrotron emission. A mixture of optically thin free-free emission and synchrotron emission has also been observed in other novae such as V445 Pup \citep{Nyamai2021} and V1535 Sco \citep{linford2017}.

\subsection{H{\sc~i} $21$-cm absorption measurements towards V3890 Sgr}

 The distance to V3890 Sgr is not well constrained, with estimates ranging from $4.4$ kpc to $9.0$ kpc using different methods \citep{schaefer2010, Munari2019b, Orio2020, 2021MNRAS.504.2122M}. %The distance estimated using \textit{Gaia} Data Release 2 (DR2; \citealp{Orio2020}) is $4.3_{-1.3}^{+2.6}$ kpc, with a huge uncertainty. As pointed out by \citet{schaefer2018distances}, the long orbital period of V3890 Sgr makes it difficult to obtain an accurate parallax measurement. Consequently, the distance to the nova determined via \textit{Gaia} parallax is unreliable. 
 Following the latest eruption, \citet{Munari2019b} determined a reddening of $E(B-V)=0.56$ mag using absorption features of optical spectral lines. Comparing this value with the three-dimensional interstellar reddening maps of \citet{Green2019} and \citet{Lallement2014}, they estimate a distance of $>4.5$ kpc. Using the surface temperature and the size of the companion star, a black body distance of $7$ kpc is derived to the nova \citep{schaefer2010}. This method relies on the orbital period of the system and assumes that the companion star fills its Roche lobe (which is far from certain in the case of a symbiotic binary).
 
\begin{figure}
  \centering
  \includegraphics[width=\columnwidth]{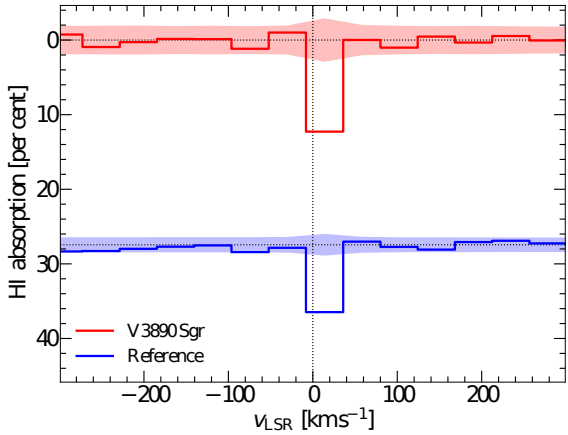}
  \caption{H{\sc~i} absorption spectrum showing an absorption between $-8~\textrm{km~s}^{-1}$ and $36~\textrm{km~s}^{-1}$ with respect to the local standard of rest. The spectrum of V3890 Sgr is plotted in red colour, and compared to an averaged spectrum of reference sources in blue colour. The shaded regions denote the noise level at $3\sigma$.}
  \label{fig:V3890Sgr_spectra_H1}
\end{figure}

 Since estimates of the distance to V3890~Sgr vary based on different observations, an attempt is made here to further constrain the distance using H{\sc~i} absorption along the line-of-sight to the nova (see \citealt{Chauhan2021} for more details about H{\sc~i} absorption with MeerKAT). %Figure~\ref{fig:V3890Sgr_schematic} shows a schematic view of the velocity field of the Milky Way and the direction of the line-of-sight of V3890 Sgr at Galactic coordinates $l=9.20^{\circ}$, $b=-06.44^{\circ}$. The schematic shows the line-of-sight through the Galactic radial velocity field, which has a tangent point (distance of greatest velocity) close to the Galactic centre.
 The epochs used to obtain the H{\sc~i} spectrum towards V3890 Sgr include radio detections of the nova when it was at its brightest ($>30$ mJy). Figure~\ref{fig:V3890Sgr_spectra_H1} shows the average MeerKAT H{\sc~i} spectrum towards V3890 Sgr, compared with the average of seven reference sources (indicated by the blue spectrum in Figure~\ref{fig:V3890Sgr_spectra_H1}) which have been offset for clarity. These reference sources are presumably background extragalactic sources, and their spectra were averaged together to yield optimal S/N. These spectra were constructed by taking an inverse-variance weighted average over spectra from seven epochs observed with MeerKAT. Line-of-sight absorption through Galactic H{\sc~i} clouds is detected towards V3890 Sgr. By identifying distinct kinematic components in the H{\sc~i} spectrum, and comparing with the spectrum of reference sources, an attempt is made to determine the distance to V3890 Sgr.
 
 %\begin{figure}
 % \centering
 % \includegraphics[width=\columnwidth]{RNV3980sgr_schematic.pdf}
 % \caption{A schematic view of V3890 Sgr sight-line through the Galactic radial velocity field plotted by Robert Hurt (Spitzer Science Center/IPAC/JPL), obtained from http://www.astro.wisc.edu/~stantzos/index.html). The blue and red contours represent radial velocity values moving towards and away from the observer respectively.}
 % \label{fig:V3890Sgr_schematic}
%\end{figure}

H{\sc~i} absorption at the level of $12\%$ is clearly detected in both the spectra of V3890 Sgr and the reference sources. In both cases, the absorption is at a velocity of $14\pm22~\textrm{km~s}^{-1}$, but is unfortunately unresolved by the data obtained in the MeerKAT 4k correlator mode. At the time of the observations, the $32$k correlator mode was not yet available. Absorption is detected at velocities lower than $36~\textrm{km~s}^{-1}$, corresponding to kinematic distances greater than $4.44^{+0.34}_{-0.31}~\textrm{kpc}$ (using a Monte Carlo technique by \citealp{Wenger2018}). This estimate is similar to the distance value obtained using %\textit{Gaia} DR2 and 
interstellar reddening as discussed earlier. Therefore, the distance constraint for this nova is consistent with being greater than $4.4$ kpc. The absolute magnitude in the $V$ filter and bolometric magnitude measurements give distance estimates of $8.75$ and $9$ kpc respectively \citep{2021MNRAS.504.2122M}. A distance of $9$ kpc is thus adopted for calculations in this work.
 
\section{Discussion}
\label{sec:synchrotron_discussion}
\subsection{Radio emission from V3890 Sgr is synchrotron dominated}
\label{subsec:synchrotron}

The radio emission of novae embedded in the winds of giant stars is dominated by synchrotron emission, as observed in systems such as RS Oph \citep{Taylor1989, Obrien2006, Rupen2008, Sokoloski2008, Eyres2009}, V745 Sco \citep{Kantharia2016} and V1535 Sco \citep{linford2017}. The non-thermal radio emission is the result of the ejecta interacting with the pre-existing circumstellar medium. A nova shock wave moving outwards populates a thin region of shocked circumstellar material with accelerated particles required for non-thermal emission. The evolution of the shock wave in symbiotic novae is similar to that of SNe following an explosion \citep[e.g.,][]{chevalier1982, Obrien2006}. The radio luminosity increases as the optically thick emitting region expands. As the emitting region expands, the free-free optical depth from the ionized red giant wind ahead of the shock decreases, and the emission becomes optically thin. 
%The fast increase in radio flux is attributed to a decrease in the optical depth of the surrounding ionized gas as the shock wave moves outward. 
The radio luminosity peaks when $\tau = 1$, and then decays as the expanding shock wave decelerates in velocity and the surrounding medium drops in density \citep{chevalier1982}.

The radio light curve of V3890 Sgr evolves through the three phases of rise, peak and decay within months following the nova eruption, and is similar to V745 Sco \citep[e.g.,][]{Kantharia2016}. The radio spectra of V3890 Sgr during the rise phase of the light curve yield a spectral index of $\alpha \approx 1.3$, consistent with optically thick emission. After the radio peak, the spectral index is steep and converges to $\alpha = -0.3$ at late times of light curve evolution (see Figure \ref{fig:radio_light_curve} and Table \ref{tab:Table_results}), an indication of optically thin synchrotron emission. Similar values of $\alpha$ are observed in sources that are strong synchrotron emission emitters such as SNe and SN remnants \citep{weiler2002, Green2019} and the helium nova V445~Pup \citep{Nyamai2021}.

\begin{figure}
  \centering
  \includegraphics[width=\columnwidth]{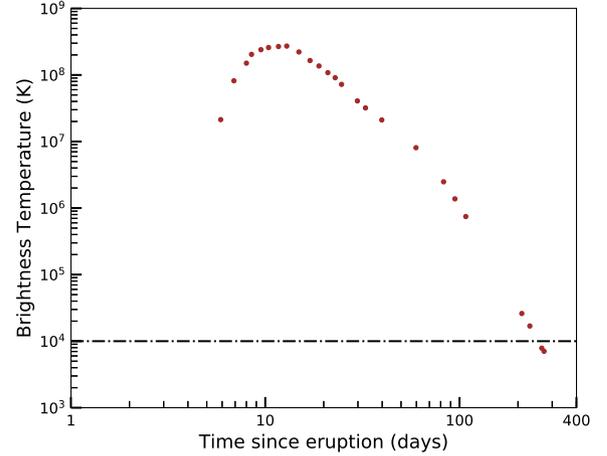}
  \caption{Brightness temperature of V3890 Sgr following the $2019$ eruption, estimated assuming a constant shock velocity of $4200$ $\textnormal {km~s}^{-1}$, and flux densities of the $1.28$ GHz observing frequency. %\textbf{The values of the brightness temperature given are lower limits given that we use a constant velocity of the shock wave. The velocity is used to set the size of the emitting region and is expected to decelerate as it interacts with the red giant wind, consequently the emitting region is expected to reduce significantly. \textcolor{red}{Redo this plot with a distance of $9$ Kpc.}}
  }
  \label{fig:V3890Sgr_T_B}
\end{figure}

To further constrain the type of emission from V3890 Sgr, the brightness temperature is determined by estimating the size of the emitting region \citep{Nyamai2021, Chomiuk2021}. To determine the angular size of the emitting region, a spherically symmetric shock wave expanding at $\approx 4200~\textnormal {km~s}^{-1}$ since $t_0$ is assumed \citep{strader2019}. Using flux densities observed at $1.28$ GHz ($23$ cm), the estimated brightness temperatures on the first $200$ days are $\gg 10^{5}$ K as shown in Figure~\ref{fig:V3890Sgr_T_B}, a strong indicator of non-thermal emission \citep[][]{Chomiuk2021}. The brightness temperature after day $200$ declines to $\approx10^{4}$ K. However, we note that the actual angular size of the emitting region may be substantially smaller than our estimated size of the emitting region, since the shock wave decelerates as it interacts with the red giant wind. The implication is that the brightness temperatures shown in Figure~\ref{fig:V3890Sgr_T_B} are lower limits. 
The spectral evolution and brightness temperatures indicate that the radio emission from V3890 Sgr is synchrotron dominated, produced in a manner similar to that in SNe, where the radio light curve also evolves on timescales of weeks to months \citep{weiler2002}.

\subsection{A model for synchrotron emission from nova blast waves}
\label{sec:synch_model}

In an environment where the nova ejecta interact with a dense surrounding medium, radio emission can be used to probe the external surrounding medium and determine its density profile, as is commonly done in radio SNe \citep[e.g.,][]{weiler2002}. The radial profile depends on the mass loss from the companion star and its shaping by binary interaction \citep[e.g.,][]{mohamed2015models,ji2013}. An interaction of the nova ejecta with the companion's wind will accelerate particles to high energies through diffusive shock acceleration \citep{Blandford1978, Bell1978} and amplify the shock magnetic field \citep{Bell2004}, producing synchrotron radiation. The interaction produces a forward shock which drives into the pre-existing circumstellar material and a reverse shock driving into the ejecta \citep{Chevalier1982b, Reynolds2017}. The radius of discontinuity $R_{\textrm {cd}}$ separates the forward- and reverse-shocked regions. The evolution of the shock fronts depends on the density structure of the nova ejecta ($\rho_{\textrm{ej}}$) and that of the surrounding medium ($\rho_{\textrm{CSM}}$; \citealt{Chevalier1982b, Tang2017}).

To interpret the radio luminosity from V3890 Sgr, this double shock system is considered. Shock wave dynamics have been observed in X-rays for recurrent novae, where the most notable characteristics of the shocked ejecta are high  temperatures, $\gtrsim10^7$ K, which decrease with time as the shock decelerates \citep[e.g.,][]{Sokoloski2006,Bode2006}. 
Immediately following its $2019$ eruption, V3890 Sgr produced hard X-rays, which were attributed to the nova outflow impacting the red-giant wind \citep{Sokolovsky2019, Orio2020, Page2020, Singh2021}. More evidence of shocks in V3890 Sgr comes from the presence of high-ionization emission lines in optical and infrared spectra \citep{Munari2019,Evans2022}.

A formalism for predicting synchrotron emission from shocks is put forward by \citet{chevalier1982}. Above a minimum energy $E_{\rm min}$, the energy spectrum of relativistic electrons can be described by a power law distribution, $N(E) = N_0~E^{-p}$ where $N_0$ is a constant, $N(E)$ is the number of particles with energy $E$, and $p$ is the power law index of the energy spectrum. The energy of a relativistic electron is $E = \gamma m_e c^2$, where $\gamma$ and $m_e$ are the Lorentz factor and mass of an electron, respectively. For the non-relativistic shock velocities observed in V3890 Sgr, we assume that $E_{\rm min}$ is the rest-mass of the electron.
The optically thin synchrotron spectrum produced by a power law distribution of electrons is also a power law, $L_\nu \propto \nu ^{-(p-1)/2}$, so that the spectral index $\alpha = -(p-1)/2$. For V3890 Sgr, 
%for the range of spectral indices, 
an average value of $\alpha \approx -0.3$ is  measured in the optically thin limit of the radio light curve. This translates to $p = 1.6$. Theoretically, in diffusive shock acceleration, the spectral index of relativistic particles $p$ is predicted to be between $2$ and $2.5$ \citep{Bell1978, Blandford1978}. However, for novae, $p$ is observed in the range of $1.2$ and $2$ \citep{Eyres2009, Finzell2018, Nyamai2021}. This could imply that relativistic electrons in novae are more evenly distributed between low and high energies, or the magnetic fields and particle density are not uniformly distributed leading to complex optical depth effects \citep{Vlasov2016}. For the non-relativistic shocks in novae, the minimum energy is taken to be the rest mass energy of the electron ($\gamma_{\rm min} = 1$;  \citealt{Chevalier1998}).

The post-shock energy density is described as $U_{\textrm {shock}} \approx \rho_{\textrm{CSM}}\, V_{\textrm {shock}}^2$, where $\rho_{\textrm {CSM}}$ is the density of the material being shocked and $V_{\textrm {shock}}$ is the velocity of the forward shock. In Chevalier's model, it is assumed that a fraction of the post-shock energy density is transferred to the accelerated electrons and the amplified magnetic field. Therefore, the energy density in relativistic electrons is $U_e =\epsilon_e\,\rho_{\textrm {CSM}}\, V_{\textrm {shock}}^2$, and the energy density in the magnetic field is $U_B =\epsilon_B\,\rho_{\textrm {CSM}}\, V_{\textrm {shock}}^2$. The efficiency factors $\epsilon_e$ and $\epsilon_B$ are used to describe the fraction of the post-shock energy in the form of relativistic electrons and amplified magnetic fields, respectively. 
%Because the shock is non-relativistic and $\gamma_{\rm min}$ is set to unity, we follow the suggestion of Harris et al.\ 2022 that only a fraction $f_{\rm nt}$ of the electrons are accelerated into the non-thermal power-law tail. This fraction can be solved for as:
%\begin{equation}
%f_\mathrm{nt} &=& 2.04 \frac{p-2}{p-1}
%        \left(\frac{\mu_e\epsilon_e}{\eta }\right) 
%        v_{\rm shock,9}^2 ~,
%\end{equation}
%where $v_{\rm shock,9}$ is in units of $10^9$ cm s$^{-1}$, $\mu_e = \rho/n_e$, and $\eta$ is the mass compression ratio (taken here to be 4 as implied by the Rankine-Hugoniot shock jump conditions). 
The normalization for the electron energy spectrum can be expressed as:
\begin{equation}
N_0 = \epsilon_e\, \rho_{\textrm{CSM}}\, V_{\textrm {shock}}^2\, (p-2)\, E_{\rm min}^{p-2}
%N_0 = 4.90 \times 10^{17} (p-1) \frac{\eta f_\mathrm{nt}}{\mu_e} \rho_{\rm CSM}
\end{equation}
in cgs units; this expression is valid for $p>2$.

\citet{Chevalier1998} expresses the flux density of synchrotron emission at frequency $\nu$ as:
 \begin{equation} 
S_\nu = 1.8~\times~10^{-22}~\bigg(\frac{\pi~R^2}{D^2}\bigg)~\frac{C_5}{C_6}~B^{-0.5}~\nu^{2.5} \Bigg(1-\textrm {exp}\bigg[-\frac{\nu}{\nu_1}^{\frac{-p-4}{2}}\bigg]\bigg)~\textrm {mJy}.
 \label{eq:opticallythinemission}
\end{equation}
Constants $C_5$ and $C_6$ are determined as a function of $p$ \citep{pacholczyk1970}. For V3890 Sgr, we take $p = 2.1$ as this modelling formalism require $p > 2.1$. Therefore, the values  $C_5 = 1.37~\times~10^{-23}$ and $C_6 = 8.61~\times~10^{-41}$ for $p=2.1$ are used here.
%$C_5 = 1.37~\times~10^{-23}$ and $C_6 = 8.61~\times~10^{-41}$
$R$ is the radius of the blast wave, and $D$ is the distance to the nova. $B$ is the post-shock magnetic field strength and is given by  $B = \sqrt{8\,\pi\,U_B}$. 
$\nu_1$ is the frequency at which the optical depth to synchrotron self-absorption (SSA) is equal to unity, given as 
\begin{equation*}
\nu_1 = 1.3~\times~10^{19}\times\bigg(\frac{4}{3}f~R~C_6~N_0\bigg)^\frac{2}{p+4}~\times~B^{(\frac{p+2}{p+4})}~\textrm{Hz}.
 \label{eq:turn_over_frequency}
\end{equation*}
The synchrotron emitting region is assumed to be between the forward shock and the contact discontinuity; its volume filling factor is quantified as $f$. %with a thickness of $10$\% of the total emitting spherical volume
We estimate the value of $f= 0.88$ using the approximations of the forward shock radius and the contact discontinuity radius given in \citet{Tang2017}. For the blast wave evolution described in \S \ref{sec:dyn}, the forward shock radius is twice that of the contact discontinuity. 
%The normalization factor yielding the number density of electrons emitting non-thermal radiation is given as
%\begin{equation}
%N_0 = \frac{\frac{\epsilon_e}{\epsilon_B} B^2(p-2)E^{p-2}}{8\pi}
%\label{eq:N0}
%\end{equation}
%where, $\propto = \frac{\epsilon_e}{\epsilon_B}$ and 
%where the definition is valid for $p>2$.

Radio synchrotron luminosity may increase at early times due to SSA or free-free absorption by the ionized gas ahead of the forward shock, depending on the amplified magnetic field, the density of the external medium and the shock wave velocity \citep{Chevalier1998, weiler1986}. It has been shown that free-free absorption is dominant for slow shock wave velocities ($\lesssim 10^4~\textrm {km~s}^{-1}$) while SSA is the dominant source of opacity for faster blast waves %($>1.1\times10^4~\textrm{km~s}^{-1}$)
\citep{Chevalier1998, Panagia2006}. 
In cases where the optical depth is dominated by an ionized wind-like medium (as described in \S\ref{sec:csm}) ahead of the shock, the free-free optical depth is defined by \citet{chevalier1981} as
\begin{align}
\begin{split}
\tau_{\textrm{ff}} &= 0.005
\bigg(\frac{{\dot M}}{10^{-6}~\textrm M_{\odot}~\textrm{yr}^{-1}}\bigg)^{2} \times
\bigg(\frac{V_\textrm {wind}}{10~\textrm{km~s}^{-1}}\bigg)^{-2} \times \bigg(\frac{\nu}{1.4~\textrm {GHz}}\bigg)^{-2}\\
&\times \bigg(\frac{R_{\textrm {shock}}}{2 \times 10^{16}~\textrm{cm}}\bigg)^{-3}
\label{eq:tau}
\end{split}
\end{align}

\subsection{The dynamics of the blast wave}\label{sec:dyn}

To predict the radio luminosity of a nova interacting with circumbinary material, we must know the radius and velocity of the blast wave. These depend on the density and velocity structure of the nova ejecta, along with the density profile of the circumbinary material.

\subsubsection{The density structure of the nova ejecta}
\label{subsubsec:Density}

The unshocked material of the nova envelope is  commonly described as expanding freely and homologously, such that $V(r) \propto r$ where $V$ is the expansion velocity %$t$ is time since eruption
and $r$ is radial distance from the white dwarf. In this case, the ejecta would show a range of velocities with the inner ejecta characterised by low velocities and the outer ejecta expanding fastest. Based on the modelling of nova remnants, the density profiles of the ejecta can be described by a power law distribution, $\rho_{\textrm{ej}} \propto r^{-n}$ with inner ejecta having $n =~2$ or $3$ and outer ejecta having $n$ between $10$ and $20$ \citep[e.g.,][]{Hauschildt1997}. Only a tiny fraction of the ejecta mass is found in the outer parts characterised by a steep power law (and this mass will be swept up very quickly, in just a few hours), so a shallow power law is adopted to describe the ejecta density profile of V3890 Sgr, such that $n~=~2$. The kinetic energy of the nova ejecta is therefore the integral of the density profile with an homologous expansion such that

%\begin{equation}
%E = \frac{M_{\textrm{ej}}}{6}\frac{V^{3}_{\textrm{max}}-V^{3}_{\textrm{min}}}{V_{\textrm{max}}-V_{\textrm{min}}}
%   \label{eq:kinetic_energy}
%\end{equation}
\begin{equation}
E = 0.17\times \bigg (\frac{M_{\textrm{ej}}}{{\textnormal M}_\odot}\bigg )\times \bigg (\frac{V^{3}_{\textrm{max}}}{{\textrm{cm~s}^{-1}}} - \frac{V^{3}_{\textrm{min}}}{{\textrm{cm~s}^{-1}}} \bigg ) / \bigg (\frac{V_{\textrm{max}}}{{\textrm{cm~s}^{-1}}} - \frac{V_{\textrm{min}}}{{\textrm{cm~s}^{-1}}} \bigg )^{-1}~\textrm{erg}
\label{eq:kinetic_energy}
\end{equation}

where $M_{\textrm{ej}}$ is the mass of the nova ejecta, $V_{\textrm{max}}$ is the maximum ejecta velocity and $V_{\textrm{min}}$ is the minimum ejecta velocity.

Observationally, the ejected masses of symbiotic recurrent novae are in the range of $10^{-7}$--$10^{-6}~{\textnormal M}_\odot$ \citep{Obrien1992, Anupama1994, Sokoloski2006, Orlando2017}. After the 1990 eruption of V3890 Sgr, the optical light curve  at $V$ band declined by three magnitude in $14$ days \citep{schaefer2010}. After the $2019$ eruption, the optical light curve also shows a fast decline (see Figure~\ref{fig:opticallightcurve}). Such a fast decline is expected for nova envelopes with mass of $< 10^{-6}~\textnormal M_\odot$ \citep{Yaron2005}. Furthermore, the same rapid decay of the optical light curve is observed in RS Oph, where its ejected nova envelope is estimated to be $(10^{-7}\leq~M_{\textrm{ej}} \leq 10^{-6})~\textnormal M_\odot$ \citep[e.g.,][]{Obrien1992, Sokoloski2006}. More evidence of a low-mass ejected envelope in V3890 Sgr is based on the fast nova evolution where high-ionization lines  appear in the spectra $\approx 18$ day following the $1990$ eruption \citep{Anupama1994}. Anupama \& Sethi use Balmer emission line fluxes to estimate the mass of the nova envelope as $\approx 10^{-7}~\textnormal M_\odot$. In our analysis we consider ejecta masses in the range $10^{-7}$--$10^{-5}~{\textnormal M}_\odot$.

\subsubsection{The density structure of the circumstellar medium}
\label{sec:csm}

The simplest model for the circumstellar material is to assume the companion star is expelling a spherically symmetric wind with constant velocity and mass-loss rate.
The density distribution of the medium is then described as 
\begin{equation}
   \rho_{\textrm {CSM}} = \frac{\dot M}{4\pi V_\textrm{wind}} r^{-2}
  \label{eq:density_wind}
\end{equation}
%= n_{\textrm{CSM}}(r)\mu m_H=
where $\dot M$ is the mass loss rate, $V_\textrm{wind}$ is the velocity of the red giant wind,  and $r$ is the distance from the companion binary.
%$n(r)$ is particle number density and $\mu~m_H$ is the mean atomic weight. 
Without prior knowledge of the distribution of circumstellar material, this spherical distribution of the red giant wind is assumed for V3890 Sgr. It is however noted that non-spherical distribution of material is common in symbiotic recurrent systems \citep{Walder2008, Booth2016}. Furthermore, the two components observed in the emission line profiles of V3890 Sgr following the 1990 eruption indicate that the nova remnant is non-spherical %asymmetric
\citep{Anupama1994}. A spherical assumption considered here is therefore a simplistic approach. 

\subsubsection{Radius and velocity of the shock front}
\label{subsubsec:radiusvelocity}

The time evolution of a shock wave propagating through the red giant wind is divided into two phases. During the first phase, when the mass of the nova envelope ($M_{\textrm{ej}}$) is much larger than that of the swept-up surrounding medium ($M_{\textrm {sw}}$), the shock is in free expansion. As the mass of the swept up material increases such that $M_{\textrm {sw}}~>M_{\textrm{ej}}$, the nova envelope enters a second phase of evolution referred to as the Sedov-Taylor phase. For spherical nova ejecta with a power-law density profile ($\rho_{\textrm{ej}} \propto r^{-2}$) interacting with circumstellar medium of density, $\rho_{\textrm{CSM}} \propto r^{-2}$, %similarity
self-similar solutions are determined for the evolution from a free expansion phase to the adiabatic expansion phase, as shown in Table $1$ of \cite{Tang2017}. For the purposes of this analysis, we assume that the red giant and white dwarf are co-located at the center of the red giant wind, a valid assumption over the course of the radio light curve because the shocked ejecta expand outside the binary system within the first day of eruption.

\citet{Tang2017} present expressions for the radius and velocity of the forward shock with time, given as $R_{\textrm{shock}} =R_{ch}\times R^*_b$,  $V_{\textrm{shock}} =V_{ch}\times V^*_b$, and $t =t_{ch}\times t^*$.
Here $R_{ch}$, $V_{ch}$, and $t_{ch}$ are the characteristic radius, velocity and time, respectively:%variables:
\begin{align}
\begin{split}
    R_{ch} & = 12.9\times\bigg(\frac{M_{\textrm{ej}}}{\textrm{M}_\odot}\bigg)\times\bigg(\frac{\overbigdot M}{10^{-5}~\textrm{M}_\odot\textrm{yr}^{-1}}\bigg)^{-1}\\
    &\times \bigg(\frac{V_\textrm{wind}}{10~\textrm{km~s}^{-1}}\bigg)~\textrm{pc}
\end{split}
\label{eq:characteristic_radius}
\end{align}

\begin{align}
\begin{split}
    t_{ch} &= 1772\times\bigg(\frac{E}{10^{51}~\textrm{erg}}\bigg)^{-0.5} \times \bigg(\frac{M_{\textrm{ej}}}{\textrm{M}_\odot}\bigg)^{1.5}\times \bigg(\frac{\overbigdot M}{10^{-5}~\textrm{M}_\odot\textrm{yr}^{-1}}\bigg)^{-1}\\
    &\bigg(\frac{V_\textrm{wind}}{10~\textrm{km~s}^{-1}}\bigg)~\textrm{yr}
\end{split}
\label{eq:characteristic_time}
\end{align}

\begin{align}
\begin{split}
    v_{ch} = 7118.24\times \bigg(\frac{E}{10^{51}~\textrm{erg}}\bigg)^{0.5} \times \bigg(\frac{ M_\textrm{ej}}{\textrm M_\odot}\bigg)^{-0.5}~\textrm{km~s}^{-1}
\end{split}
\label{eq:characteristic_velocity}
\end{align}

%$E$ is the kinetic energy of the nova ejecta.
$R^*_b$, $V^*_b$ are analytical dimensionless quantities defined by \citet{Tang2017}, and can be written as a function of $t^{*} = t/t_{ch}$ as
\begin{align}
%\begin{split}
    R^*_b = 2.91489\times t^{*}(1 + 9.36843t^{* 0.567})^{-0.5882}
   % V^*_b = \frac{dR^*}{dt^*} = (2.91489~+~18.2t^{*0.567})(1~+~9.36843t^{*0.567})^{-1.5882}
%\end{split}
\label{eq:dimensionless_equations}
\end{align}
\begin{align}
 V^*_b = \frac{dR^*}{dt^*} = (2.91489~+~18.2t^{*0.567})(1~+~9.36843t^{*0.567})^{-1.5882.}
\end{align}

\begin{figure}
  \centering
  \includegraphics[width=\columnwidth]{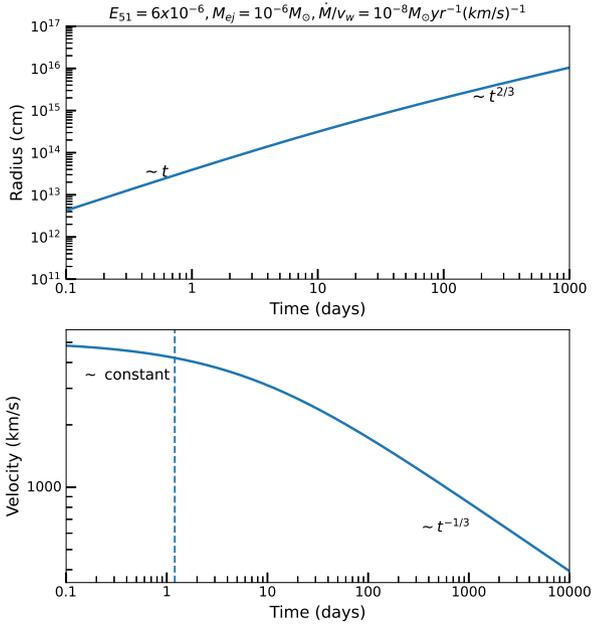}
  \caption{Top: Radial evolution for a nova forward shock interacting with a circumstellar material, assuming ejecta mass of 10$^{-6}$ M$_{\odot}$, kinetic energy $E= 6 \times 10^{45}$ erg, and $\dot{M} = 10^{-7}$ M$_{\odot}$ yr$^{-1}$ for $V_{\rm wind}$ = 10 km s$^{-1}$. This represents the free expansion ($R_{\rm shock} \propto t$) and the Sedov-Taylor phase of the evolution ($R_{\rm shock} \propto t^{2/3}$). Bottom: Velocity evolution of the forward shock as it interacts with the surrounding medium. At the beginning of the evolution, during the free expansion phase, the velocity is constant. Later, the shock enters the Sedov-Taylor phase of evolution and $V_{\textrm {shock}} \propto t^{-1/3}$. The vertical dashed line represents the initial maximum velocity of $\approx4200~\textrm{km~s}^{-1}$ of the nova ejecta on day $1.2$ after the nova eruption.}
  \label{fig:Radius_Velocity_evolution}
\end{figure}

The predicted radius ($R_{\textrm {shock}}$) and velocity ($V_{\textrm {shock}}$) of the blast wave are shown in Figure~\ref{fig:Radius_Velocity_evolution}. The radius and velocity of the shock depend on the ejecta mass ($\approx10^{-5}-10^{-7}~\textrm{M}_\odot$) and kinetic energy of the nova ejecta, along with the density of the circumbinary material (i.e., ${\dot M}/ V_\textrm{wind}$). The shock radius first grows linearly with $t$ and later as $t^{0.67}$. Similarly, the shock wave velocity first remains at a near-constant value but decreases as $t^{-0.33}$ at later times. The flux densities depend on these quantities, but also on the distance to the nova, $\epsilon_e$, and $\epsilon_B$. We tweaked $M_{\textrm{ej}}/ E$ to yield maximum expansion velocities at early times (i.e., during free expansion; Figure \ref{fig:Radius_Velocity_evolution}) consistent with the maximum ejecta velocity of $\approx4200~\textrm{km~s}^{-1}$, as estimated from optical spectroscopy \citep{strader2019}.

%\begin{figure}
 % \centering
 % \includegraphics[width=\columnwidth]{Radius_evolution.eps}
%  \caption{Radial evolution for a nova forward shock interacting with a circumstellar material. This represents the free expansion ($R \propto t$) and the Sedov-Taylor phase of the evolution ($R \propto t^{2/3}$).}
 % \label{fig:Radius_evolution}
%\end{figure}

%\afterpage{\clearpage}
%\begin{figure}
 % \centering
  %\includegraphics[width=\columnwidth]{Velocity_evolution.eps}
 % \caption{Velocity evolution of the forward shock as it interacts with the surrounding medium. At the beginning of the evolution, during the free expansion phase, the velocity is constant. Later, the shock enters the Sedov-Taylor phase of evolution and $v_s \propto t^{-1/3}$.}
 % \label{fig:Velocity_evolution}
%\end{figure}

\subsection{Modelling the radio light curve of V3890 Sgr}
\label{subsec:modelling}

\subsubsection{Constraining the mass of the ejected envelope}

The radio light curve of V3890 Sgr is compared to the model used to explain non-thermal emission in supernova explosions \citep[\S \ref{sec:synch_model}; ][]{Chevalier1998}. %All the model input parameters are held constant except 
The model depends on various input parameters, including: 
$p=2.1$, $V_{\textrm{wind}} = 10~\textnormal {km~s}^{-1}$, and the distance to the nova ($9$ kpc). The velocity and radial profile of the expanding material are as discussed in \S \ref{subsubsec:radiusvelocity}.

\begin{figure*}
\minipage{0.33\textwidth}
  \includegraphics[width=\linewidth]{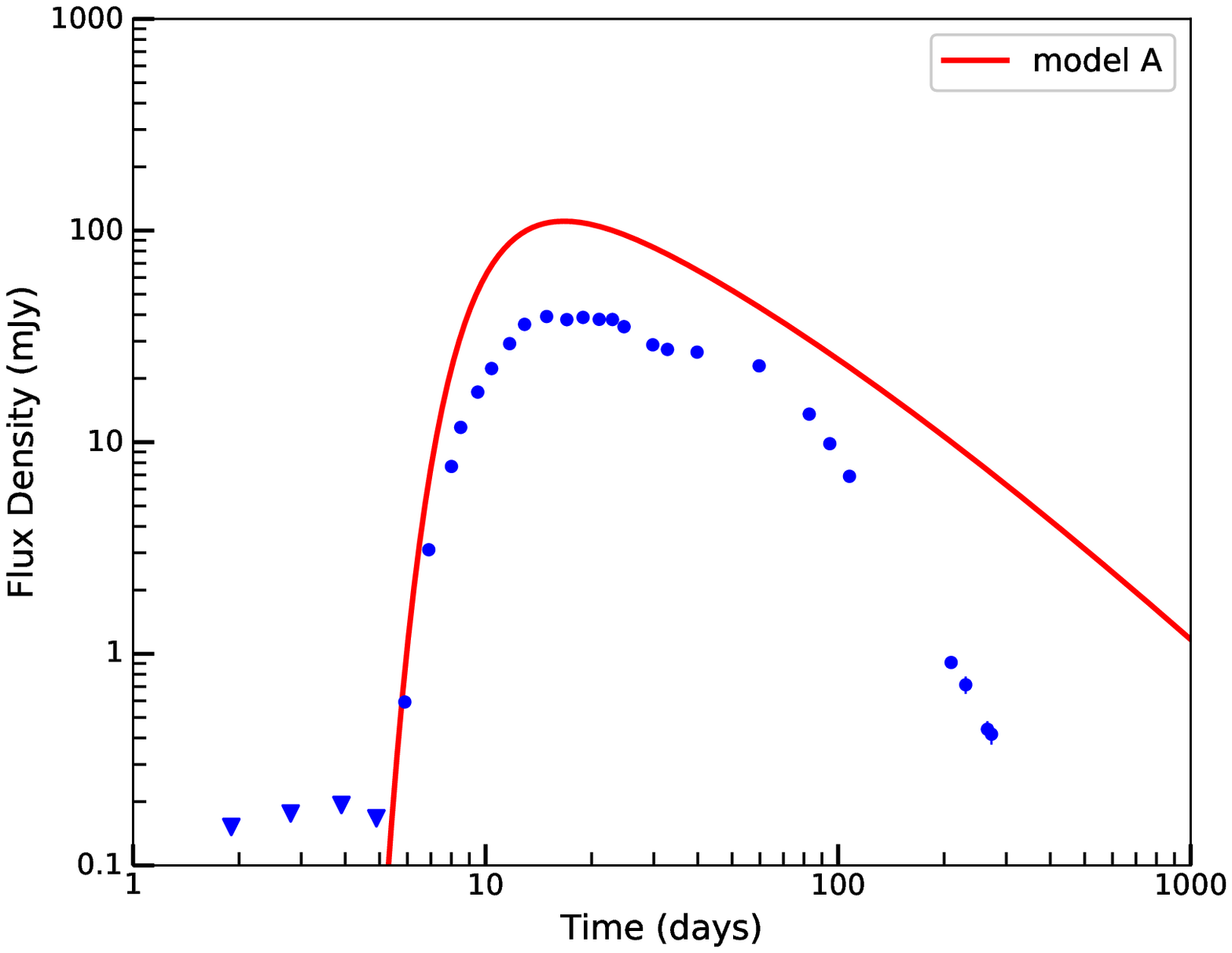}
  %\caption{(a)}
  %\label{fig:awesome_image1}
\endminipage\hfill
\minipage{0.33\textwidth}
  \includegraphics[width=\linewidth]{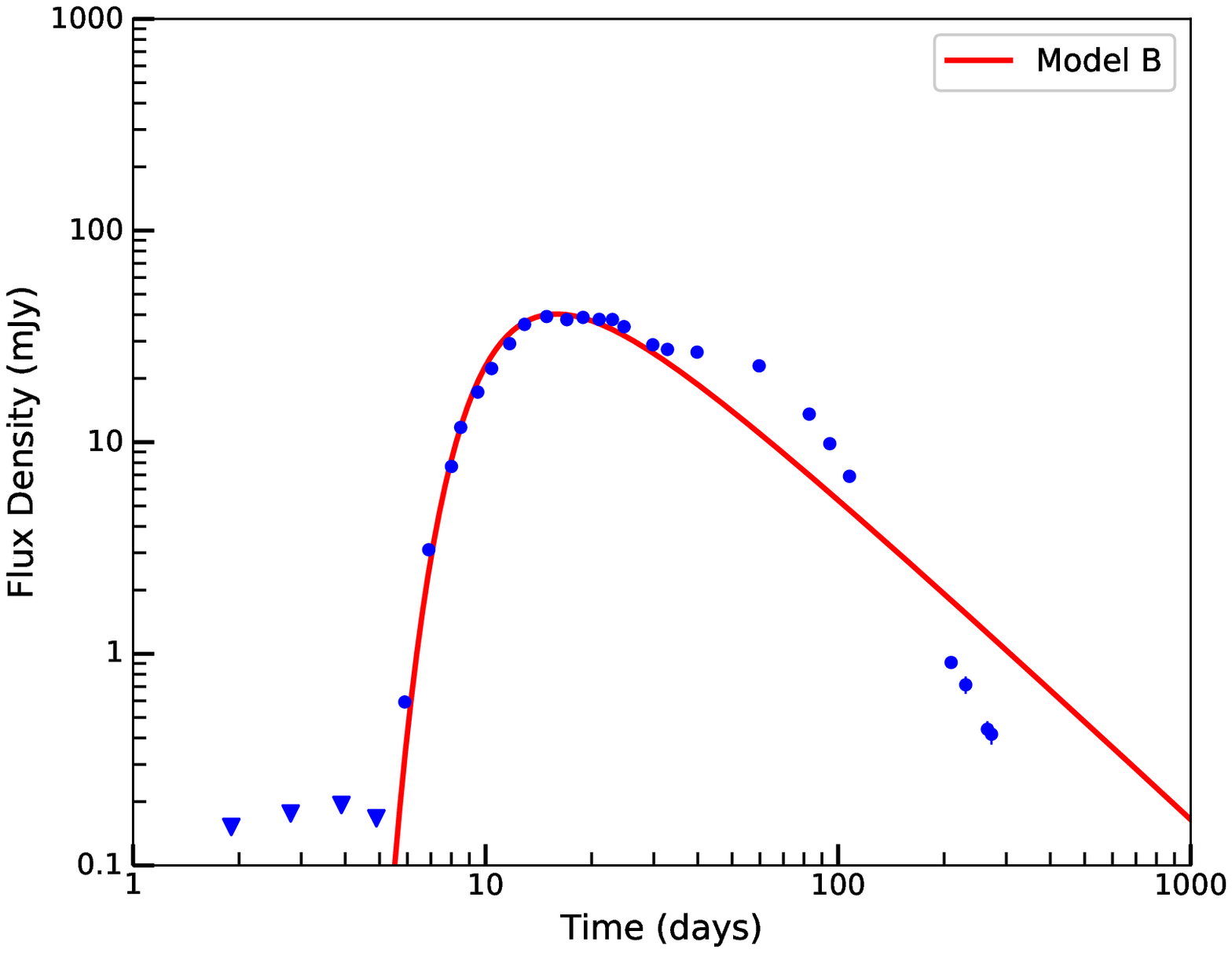}
  %\caption{(b)}
  %\label{fig:awesome_image2}
\endminipage\hfill
\minipage{0.33\textwidth}%
  \includegraphics[width=\linewidth]{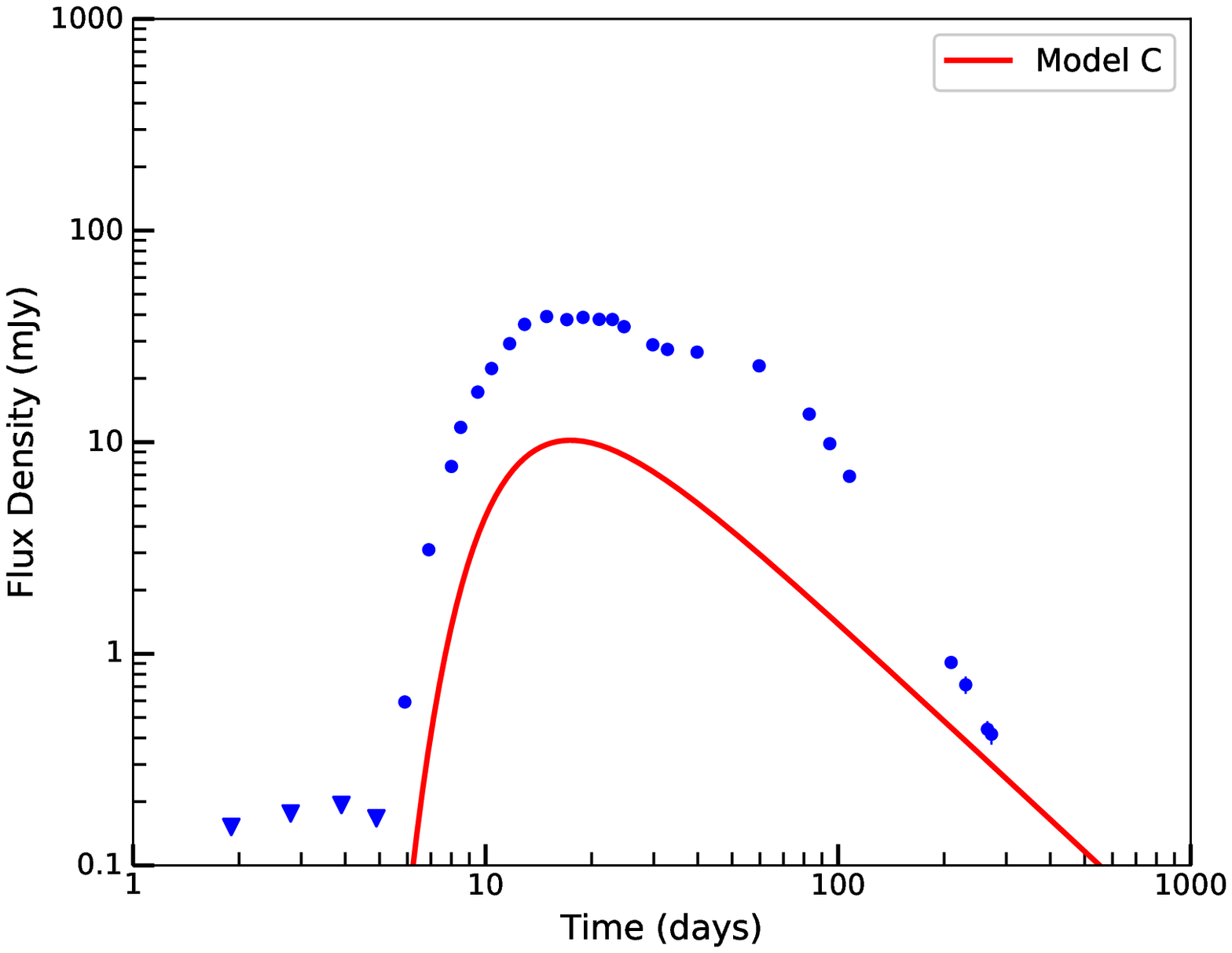}
  %\caption{c}
  %\label{fig:awesome_image3}
\endminipage
\caption{Models of radio emission produced by synchrotron emission undergoing free-free absorption are represented as red solid lines and superimposed on the MeerKAT $1.28$ GHz light curve of V3890 Sgr (blue points). The three panels
%, left, centre and right 
correspond to different models as described in Table~\ref{tab:aconf}. For a particular value of $M_{\textrm{ej}}$, the kinetic energy is varied to yield a maximum velocity on day $1.2$.
%The ratio of $M_{\textrm{ej}}/ E$ is held constant to yield a maximum velocity. 
For model A (left panel) with $M_{\textrm{ej}}= 10^{-5}~\textnormal M_\odot$, the synchrotron emission is significantly more luminous and longer lasting than the observations. The middle plot shows model B for $M_{\textrm{ej}}= 10^{-6}~\textnormal M_\odot$, and is a good match to the observations during the rise and peak of the radio emission. The right plot represents model C for $M_{\textrm{ej}}= 10^{-7}~\textnormal M_\odot$, and is not luminous enough to match the observations.}
\label{fig:modellightcurve1}
\end{figure*}

 First, we consider a range of $M_{\textrm{ej}}$ between $M_{\textrm{ej}}= 10^{-5}~\textnormal M_\odot$ and $10^{-7}~\textnormal M_\odot$.  Figure~\ref{fig:modellightcurve1} shows model radio light curves for three different ejecta masses: $M_{\textrm{ej}}= 10^{-5}~\textnormal M_\odot$, $10^{-6}~\textnormal M_\odot$, and $10^{-7}~\textnormal M_\odot$. In all three cases, we assume ${\dot M}= 3.2~\times~10^{-8}~\textrm M_\odot~\textrm{yr}^{-1}$ for $V_\textrm{wind} = 10~\textrm{km~s}^{-1}$ and $\epsilon_e = \epsilon_B = 0.01$. The value of ${\dot M}$ is estimated based on the fit of the radio fluxes peak.
 % (see Table \ref{tab:aconf}).
 %We also hold $M_{\textrm{ej}}/ E$ constant to yield a maximum velocity of
 For a particular value of $M_{\textrm{ej}}$, the kinetic energy is varied to yield a maximum velocity of $\approx 4200$ km s$^{-1}$ on day $1.2$ \citep{strader2019}. %A higher value of $M_{\textrm{ej}}/ E$ causes the light curve to become brighter and the radio emission to last longer (Model A, the left panel of Figure~\ref{fig:modellightcurve1}).
 
 For model A (left panel of Figure~\ref{fig:modellightcurve1}), the synchrotron emission is significantly more luminous and longer lasting than the observations. With the parameters given in model B in Table \ref{tab:aconf},  $M_{\textrm{ej}}= 10^{-6}~\textnormal M_\odot$ agrees well with the data during the rise and peak of the radio emission (middle panel of Figure~\ref{fig:modellightcurve1}). For model C, the right panel of Figure~\ref{fig:modellightcurve1}, the radio light curve is less bright and the radio emission does not last very long. None of the models predict the decay phase of the radio light curve accurately. A discussion of why this is the case is included in \S \ref{subsubsec:curve peak}.
 
 %\begin{landscape}
%\end{landscape}\\

\begin{table*}
    \centering
    \caption{Model parameters.}
    \label{tab:aconf}
\begin{tabular}{lcccccr}
\hline
Model& Frequency & Kinetic energy & Ejecta mass  & $\epsilon_e$ & $\epsilon_B$ & RG wind $\dot{M}$ \\
& (GHz) &  (erg) & ($\textnormal M_\odot$) &  &  & ($\textnormal{M}_\odot~\textnormal{yr}^{-1}$)\\
\hline 
A & 1.28 &$4.5\times10^{44}$  & $10^{-5}$ & 0.01 & 0.01 & $3.2\times10^{-8}$ \\
B & 1.28 &$6.0\times10^{43}$  & $10^{-6}$ & 0.01 & 0.01 & $3.2\times10^{-8}$ \\
C &1.28 & $1.4\times10^{43}$  & $10^{-7}$ & 0.01 & 0.01  & $3.2\times10^{-8}$ \\
D &1.28 & $4.5\times10^{44}$ & $10^{-5}$ & 0.004 & 0.004 & $3.2\times10^{-8}$ \\
E & 1.28 & $1.4\times10^{43}$  & $10^{-7}$ & 0.01 & 0.1  & $3.2\times10^{-8}$ \\
F & 1.78 & $6.0\times10^{43}$  & $10^{-6}$ & 0.01 & 0.01  & $3.2\times10^{-8}$ \\
G & 5.0 & $6.0\times10^{43}$  & $10^{-6}$ & 0.01 & 0.01  & $3.2\times10^{-8}$ \\
H & 7.0 & $6.0\times10^{43}$  & $10^{-6}$ & 0.01 & 0.01 & $3.2\times10^{-8}$ \\
\hline 
\end{tabular}
\begin{tablenotes}
\item Here we assume a wind velocity, $V_\textrm{wind} = 10~\textrm{km~s}^{-1}$.
\end{tablenotes}
\end{table*}

\begin{figure}
  \centering
  \includegraphics[width=\columnwidth]{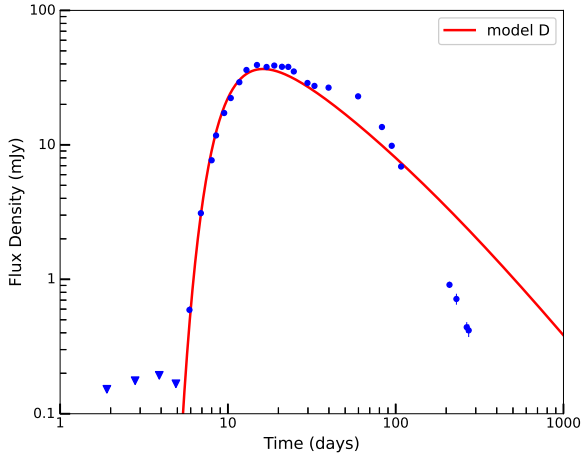}
  \caption{Model of radio emission produced by synchrotron emission undergoing free-free absorption (represented as the red solid line), superimposed on the MeerKAT light curve (blue points) for a massive ejection  $M_{\textrm{ej}}= 10^{-5}~\textnormal M_\odot$ with a shock that is less efficient at accelerating electrons and amplifying magnetic fields; see model D in Table \ref{tab:aconf} for details.} %and $\epsilon_e = \epsilon_B = 0.004$. The other model parameters are as described by Model A in Table \ref{tab:aconf}, equipartition of energy such that $\epsilon_e = \epsilon_B < 0.004$ and ${\dot M}= 4.2 \times 10^{-8}~\textrm M_\odot~\textrm{yr}^{-1}$ for $V_\textrm{wind} = 10~\textrm{km~s}^{-1}$.}
  \label{fig:modellightcurve_massive_ejection}
\end{figure}

 For the more massive ejection ($M_{\textrm{ej}}= 10^{-5}~\textnormal M_\odot$) to not overpredict the radio flux densities, the model would require less energy in accelerated particles and amplified magnetic fields, such that $\epsilon_e = \epsilon_B = 0.004$. This is consistent with estimates of particle acceleration implied by modelling of $\gamma$-ray emission from a symbiotic nova \citep[e.g.,][]{Abdo2010}. Additionally, \citet{sarbadhicary2017} have also shown $\epsilon_e$ to be lower than $0.01$ in non-relativistic shocks. Adopting these values of $\epsilon_e$ and $\epsilon_B$ produces a model D shown in Figure~\ref{fig:modellightcurve_massive_ejection} and tabulated in Table \ref{tab:aconf}.

\begin{figure}
  \centering
  \includegraphics[width=\columnwidth]{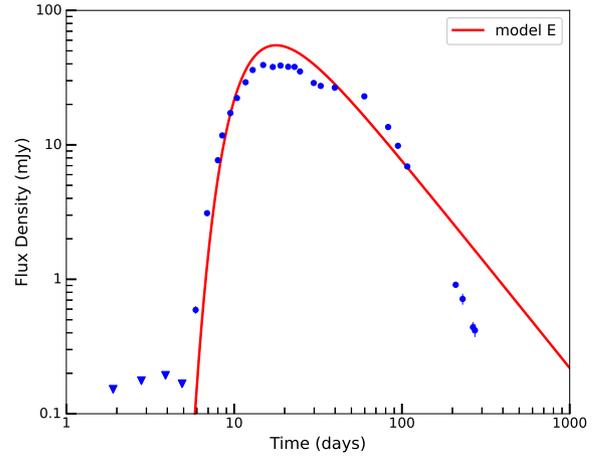}
  \caption{Model of radio emission produced by synchrotron emission undergoing free-free absorption represented (red solid line) superimposed on the MeerKAT light curve (blue points) for a less massive ejection $M_{\textrm{ej}}= 10^{-7}~\textnormal M_\odot$ that is more efficient at amplifying the post-shock magnetic field; see model E in Table \ref{tab:aconf} for details.}
  % $\epsilon_e = 0.01$, and $\epsilon_B = 0.1$. Other model parameters are describe as Model C in Table \ref{tab:aconf}.}
  %with ${\dot M}= 2.0 \times 10^{-8}~\textrm M_\odot~\textrm{yr}^{-1}$ for $V_\textrm{wind} = 10~\textrm{km~s}^{-1}$.}
  \label{fig:modellightcurve_lessmass_ejection}
\end{figure}

For the less massive ejection ($M_{\textrm{ej}}= 10^{-7}~\textnormal M_\odot$) to not underpredict the radio flux densities, the model requires a high efficiency of either accelerating particles or  amplifying magnetic fields, if not both. As $\epsilon_B$ is the more poorly constrained parameter \citep[e.g.,][]{chevalier2006,Lundqvist20}, we let it vary while holding $\epsilon_e$ fixed, finding a good match to the observed light curve with $\epsilon_e = 0.01$ and $\epsilon_B = 0.1$ (Figure \ref{fig:modellightcurve_lessmass_ejection}; Model E in Table \ref{tab:aconf}). This produces a better fitting model compared to model C. %${\dot M}= 3.2 \times 10^{-8}~\textrm M_\odot~\textrm{yr}^{-1}$ for $V_\textrm{wind} = 10~\textrm{km~s}^{-1}$.
However, since $\epsilon_B$ has been shown to vary significantly with values as low as $3~\times~10^{-3}$ \citep{Lundqvist20}, this less massive ejection model should be considered with caution. 

The most likely mass of the ejecta envelope is therefore, between $10^{-5}~\textnormal M_\odot$ and $10^{-6}~\textnormal M_\odot$. These values of ejecta mass are similar to those estimated in recurrent novae T Pyx \citep{2014Nelson} and RS Oph \citep{2022Pandey}.

 %Since the more massive ejecta of $M_{\textrm{ej}}= 10^{-5}~\textnormal M_\odot$ is inconsistent with the fast decay of the optical light curve and theoretical models as explained in \S \ref{subsubsec:Density}, the ejecta mass of V3890 Sgr following the $2019$ eruption is likely to be $M_{\textrm{ej}}= 10^{-6}~\textnormal M_\odot$.  

\begin{figure*}
\minipage{0.33\textwidth}
  \includegraphics[width=\linewidth]{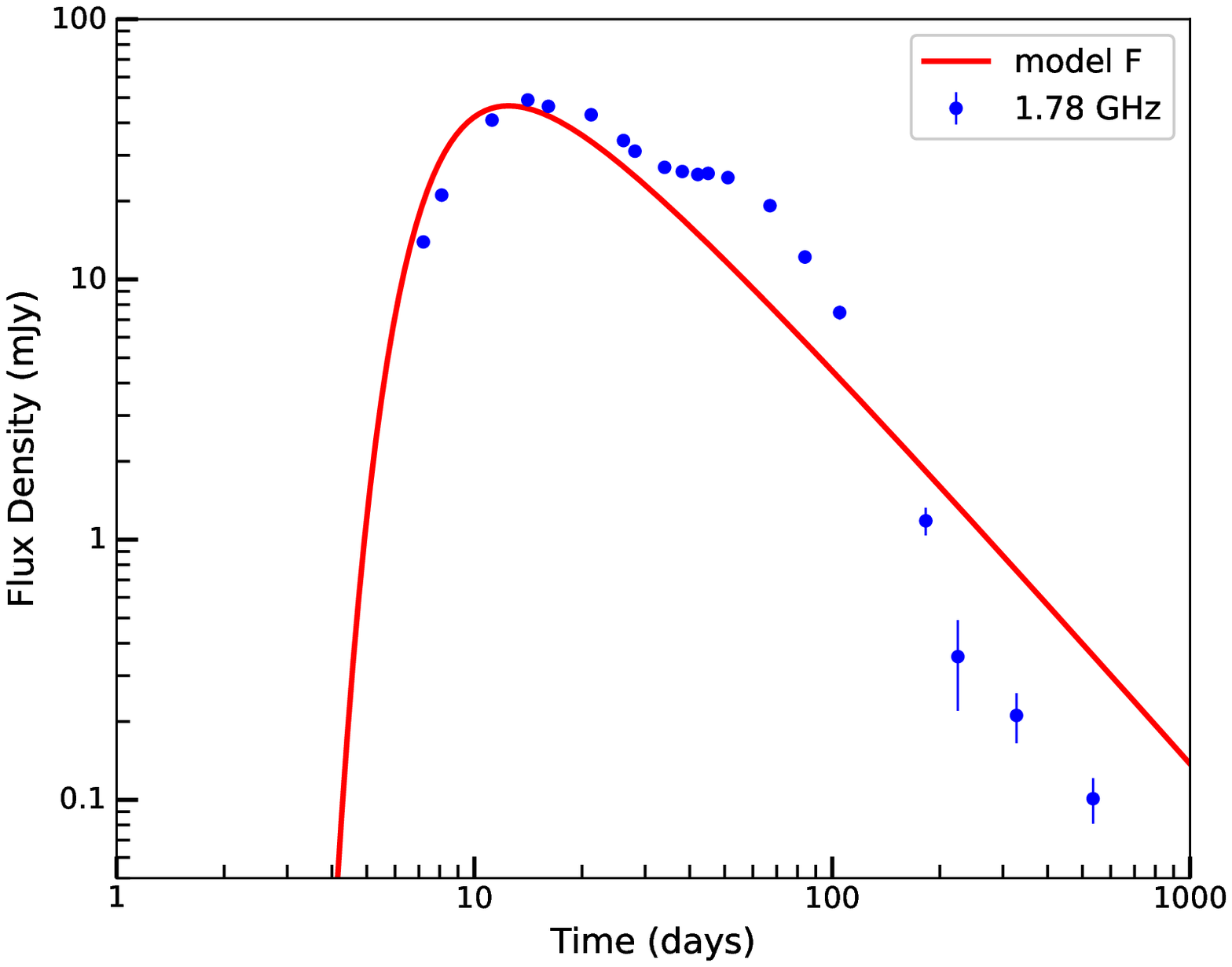}
  %\caption{(a)}
  %\label{fig:awesome_image1}
\endminipage\hfill
\minipage{0.33\textwidth}
  \includegraphics[width=\linewidth]{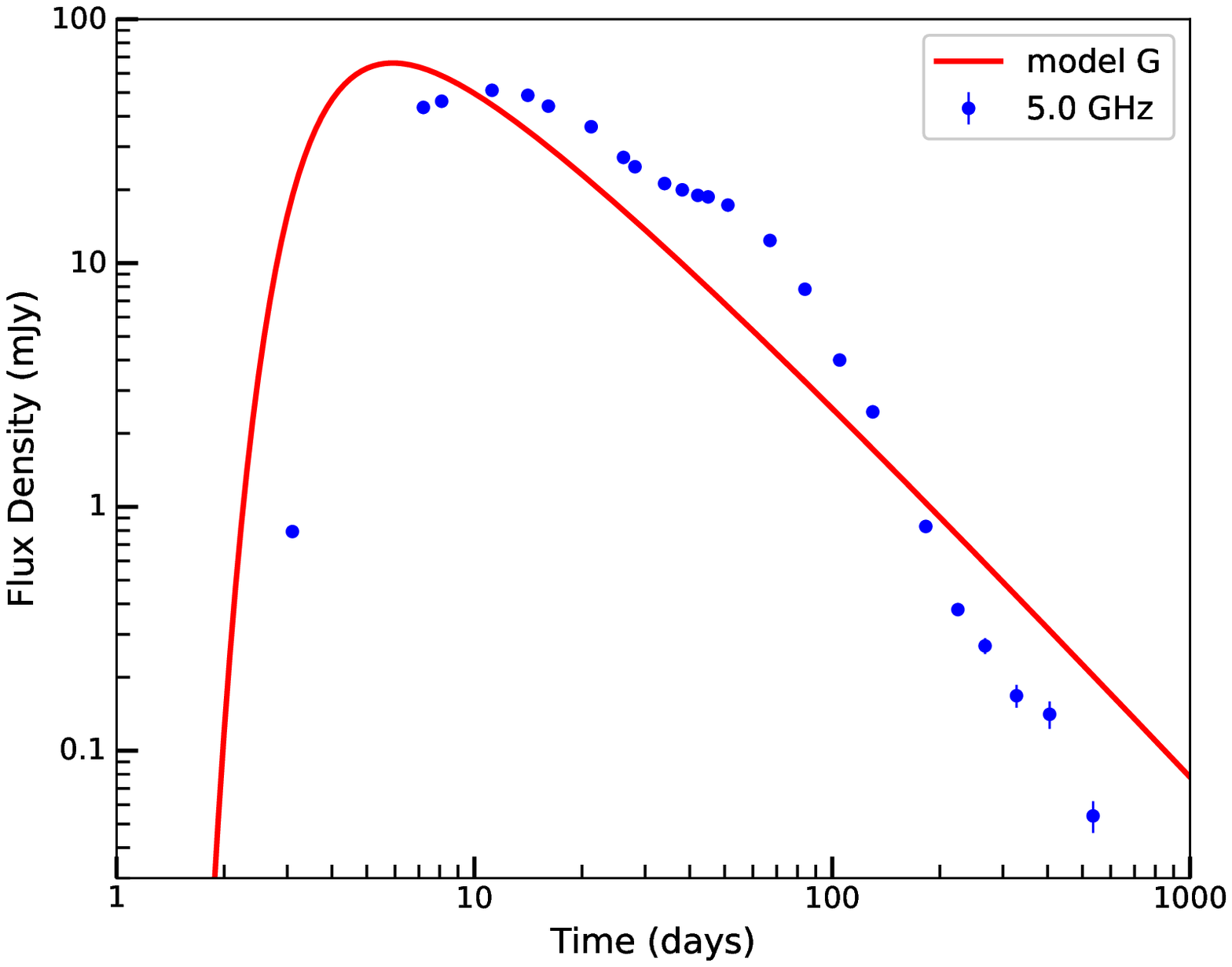}
  %\caption{(b)}
  %\label{fig:awesome_image2}
\endminipage\hfill
\minipage{0.33\textwidth}%
  \includegraphics[width=\linewidth]{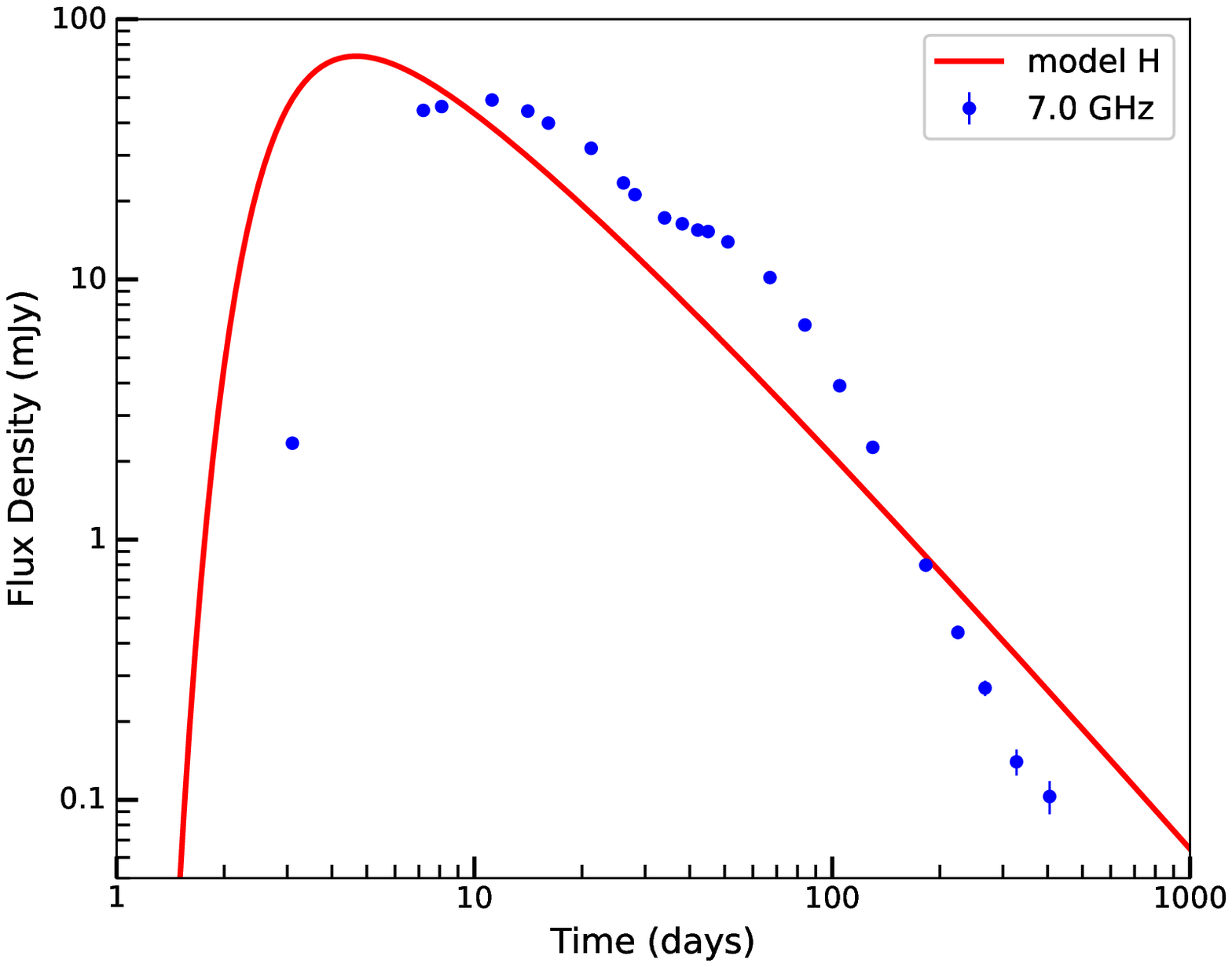}
  %\caption{c}
  %\label{fig:awesome_image3}
\endminipage
\caption{Models of radio emission produced by synchrotron emission undergoing free-free absorption are represented as red solid lines and superimposed on the VLA data (blue points) observed at frequencies $1.78$ GHz (left panel), $5.0$ GHz (middle) and $7.0$ GHz (right). The three panels from left to right represent models F, G, and H respectively (see Table \ref{tab:aconf}).}
\label{fig:modellightcurve_VLA}
\end{figure*}

%\subsubsection{Constraining the mass-loss rate of the red giant}
%An additional, interesting constraint on the density profile at small radius is the turn-on of radio emission at other frequencies. Basically--at the higher VLA frequencies, the synchrotron emission becomes transparent faster, and therefore contains information about what is going on at smaller radius. i think we should compare our model of the first radio peak with our *multi-frequency* light curve, not just the light curve at 1.4 ghz.
The model of the first radio peak is compared to the VLA radio light curves at frequencies between $1.78$ GHz and $7$ GHz as shown in Figure~\ref{fig:modellightcurve_VLA}.

 %\begin{landscape}

At higher frequencies, the ejected material becomes optically thin faster for an homologous expanding material and thus provides information at smaller radii. By keeping the ejected mass and ${\dot M}$ at constant values of $10^{-6}~\textnormal M_\odot$ and ${\dot M}= 3.2~\times 10^{-8}~\textrm M_\odot~\textrm{yr}^{-1}$, the MeerKAT model does not provide proper fits for any of the VLA radio light curves (see Figure \ref{fig:modellightcurve_VLA}). This implies that the CSM profile cannot be described with $\rho_{\textrm CSM} \propto r^{-2}$ at the regions closest to the compact object.

%the VLA radio light curve peaks provide an estimate of ${\dot M}= 3.5~\times 10^{-8}$ to $6.0~\times 10^{-8}~\textrm M_\odot~\textrm{yr}^{-1}$ for $V_\textrm{wind} = 10~\textrm{km~s}^{-1}$ (models F, G, and H in Table \ref{tab:aconf}).

Even though the prediction of synchrotron luminosity is sensitive to the assumptions made, in all the cases presented above, the radio light curve provides an estimate of the mass-loss rate of the red-giant companion of $3.2\times 10^{-8}~\textnormal M_\odot~\textrm{yr}^{-1}$. Using radio observations and assuming spherically distributed ionized circumstellar material in symbiotic systems, \citet{Seaquist1990} determined $\overbigdot M_{\textrm wind} = 10^{-7} \textnormal {~M}_\odot~\textnormal {yr}^{-1} $ for most red giant stars assuming $V_\textrm{wind}$= $10~\textrm{km~s}^{-1}$. However, studies of the most studied recurrent nova RS Oph imply a mass loss rate of $10^{-8}~\textnormal M_\odot~\textrm{yr}^{-1}$ \citep{Vaytet2007, vanLoon2008}. The mass-loss rate of V3890 Sgr estimated from its radio light curve  is in the range of those found for red giants in symbiotic systems and RS Oph.
    
\subsubsection{Explaining the second radio light curve peak}
\label{subsubsec:curve peak}

While our simple model of a $\rho_{\textrm CSM} \propto r^{-2}$ circumstellar medium fits the radio light curve rise and peak of V3890 Sgr well, the light curve then plateaus around days $\sim$30--60, in excess of the model (Figures \ref{fig:modellightcurve1}--\ref{fig:modellightcurve_VLA}; see also Figure \ref{fig:radio_light_curve}). We call this excess emission the ``second peak'' of the radio light curve. Before day $17$, the radio flux is rising. During the decay phase of the radio light curve of V3890 Sgr, the flux densities can be described by a power law relationship with time such that $S_\nu \propto~t^\beta$. Around day $17$, after the first radio peak, the radio emission flattens such that $\beta = -0.5$. After the second peak, from day $60$ on, the radio flux density rapidly declines as $\beta = -2.8$, which is a steeper decay than predicted by our model with a $\rho_{\textrm CSM} \propto r^{-2}$ circumstellar medium. This standard model is described by $\beta = -1.3$ for the decline phase \citep{chevalier2006}. %An optically thin phase rate of decline similar to that of V3890 Sgr is observed in the radio light curve of Rs Oph \citep{Kantharia2007}.

The synchrotron luminosity produced when a nova remnant interacts with the circumstellar medium is expected to rise, peak and decay in a span of a few months \citep{Kantharia2012}. The radio luminosity in this framework decays during the optically thin phase of the radio light curve due to a decrease in circumbinary density and/or shock wave velocity \citep{weiler2002}. This second peak cannot be explained by a single interaction with a spherically distributed wind-like circumstellar medium.

Double-peaked radio light curves have been observed previously in novae with giant companions, but on different timescales. For example the radio light curve of the $2006$ eruption of nova RS Oph showed a first peak around day $8$, which was followed by a second peak on day $40$ \citep{Eyres2009}. Based on its spectral evolution, the authors concluded that the radio emission from the nova is mixed thermal and non-thermal radiation. V1535 Sco is another symbiotic nova where the radio light curve shows several radio peaks with the first one recorded on day $25$ \citep{linford2017}. The spectral evolution of V1535 Sco is also consistent with a mixture of thermal and non-thermal emission \citep{linford2017}.

V3890 Sgr is the first symbiotic recurrent nova where the second radio bump is clearly dominated by non-thermal emission, given that the spectral index is predominantly negative with values of $\alpha$ ranging between $-0.3$ and $-0.4$, see Table \ref{tab:Table_results}. The brightness temperature also remains high ($\gtrsim 10^5$ K) even at day 100 (Figure \ref{fig:V3890Sgr_T_B}).
% (see \S \ref{subsec:synchrotron}). 
%compared to that observed in SNe \citep{chevalier2006}. 

 So what might explain the second radio peak and the subsequent fast decay of V3890 Sgr? The luminosity remaining bright compared to the model translates to either an excess of shock velocity or an excess
 %an increase in the velocity of the shock wave or an increase in the 
 density of material being shocked, relative to our simple model.
 %as established in the helium nova V445 Pup for example \citep{Nyamai2021}. 
 While many novae show multiple ejections and increases in ejecta velocity as an eruption proceeds \citep{Aydi+20}, the fast eruptions of symbiotic recurrent novae can generally be described as a single ejection decelerating as it sweeps up circumstellar material \citep[e.g.,][]{Walder2008, Orlando+09, Munari+11}. We therefore interpret V3890 Sgr's deviations from our simple model as complex structure in the circumstellar material (i.e., the density distribution does not just decrease as $r^{-2}$, as expected if the red giant companion powered a wind with constant velocity and mass loss rate). 

 In fact, it has been shown that the surrounding medium in systems like V3890 Sgr is likely to be structured and aspherical. %Simulations of interacting components in a binary system show the structure of circumstellar material.
 For example, simulations of RS Oph show that material lost from the companion star ends up concentrated in the orbital plane during the mass transfer process \citep{Walder2008, Booth+16}. Mass loss from the outer Lagrangian points imposes spiral waves on circumstellar material and can create complex structure as the spirals interact.  In addition, a non-uniform distribution of material may arise due to changes in the mass-loss rate or wind velocity from the companion star. Finally, recurrent nova eruptions likely sweep up circumstellar material, leading to shells bounded by lower density cavities \citep{Moore_Bildsten12, Darnley2019}. 
 %This concept has been investigated in the nova M31N 2008-12a by \citet{Darnley2019}.

 %The exception to this scenario would be that the ejected envelope was travelling in a non-uniformly distributed medium and encountered denser medium which slowed it down or the eruption was less energetic and did not travel further than $\sim 10^{15}$ cm. There are no radio observations obtained during the $1992$ eruption and therefore it is not possible to constrain this scenario.
 
The excess in radio emission between days $\sim$30--60 therefore implies a relatively dense ``shell" of CSM at $\sim 10^{15}$cm from the binary (see the radius evolution in Figure \ref{fig:Radius_Velocity_evolution}). The rapid decline of the radio light curve after day $60$ implies circumstellar material whose density declines more steeply than $\propto r^{-2}$ (see Equation~(3) in \citealp{Nyamai2021}), which can be pictured as a lower density region at radii $\gtrsim 1.3 \times 10^{15}$ cm.  Similar low density material at $\approx10^{15}$ cm is implied in another symbiotic nova V407 Cyg \citep{Chomiuk2012}. While such a shell/cavity structure is tempting to blame on past nova eruptions sweeping up the giant's wind, this is unlikely since the material ejected during V3890 Sgr's $1992$ eruption should be at a radii of $\sim 10^{16} \textrm{cm}$ based on Figure~\ref{fig:Radius_Velocity_evolution}. Unless the $1992$ eruption was quite different from the 2019 eruption (contrary to the findings of \citealt{schaefer2010}, who find that eruptions in the same recurrent nova system generally match each other well), it is difficult to imagine how the $2019$ ejecta could ``catch up'' to the $1992$ ejecta. The most likely cause of the structure in the circumstellar material is therefore the mass transfer/accretion process itself \citep{Mohamed_Podsiadlowski07, Walder2008}.

Radio observations are not the only evidence for complex structure in the circumstellar material around V3890 Sgr. \emph{Chandra} High Energy Transmission Grating observations of V3890~Sgr on day $7$ showed X-ray emission lines that are asymmetric, blue-shifted and indicating multiple plasma temperatures \citep{Orio2020}. Orio et al.\ suggest that a non-uniform distribution of the circumbinary medium could be the source of the different emission region temperatures. The H$\alpha$ emission line of V3890 Sgr obtained using the Asiago $1.22$m telescope does not show a change in the width, days following the $2019$ eruption \citep{Munari2019atel}. This implies that there is either no or minimal deceleration of the novae ejecta. Evolution of emission lines at infrared wavelengths indicate both fast uninterrupted  polar outflow and a slow equatorial outflow as a result of encounter with surrounding medium \citep{Evans2022}. Binary interaction simulations of systems similar to V3890 Sgr reveal asymmetries in the distribution of circumbinary material, such that the dense material is concentrated at the orbital plane and the less dense material is concentrated at the polar directions of the binary systems \citep{Booth+16}. The radio emission is therefore possibly originating from the interaction with dense material while the less dense material expands without much deceleration.

 \subsection{V3890 Sgr as a progenitor of SNe Ia}

The estimated low ejecta mass for the nova envelope (in the range of $10^{-7}$--$10^{-6}~{\textnormal M}_\odot$) and the short recurrence time of $28$ years imply that V3890~Sgr hosts a massive white dwarf ($\gtrsim 1.2~\textrm{M}_\odot$, \citealp{Yaron2005}), making it an interesting candidate for a SN Ia progenitor system. Non-detections of radio emission from SNe Ia have provided an opportunity to rule out candidate single degenerate progenitor systems through comparison of radio models and radio luminosity upper limits \citep[e.g.,][]{Panagia2006,Chomiuk2016,Lundqvist20}.
The radio observations of SNe Ia that are used to rule out symbiotic systems are typically obtained a few weeks to months after explosion. However, in our analysis of V3890 Sgr, we detect circumstellar material at radii $\lesssim$ few $\times 10^{15}$ cm, with a rapid fall-off in density at larger radii. Such CSM would therefore be best constrained if radio observations of SNe Ia were obtained within a $\sim$day following explosion (see Equation 10 of \citealt{Chomiuk2016}). 

The only SN Ia with such early radio follow-up published is SN\,2011fe, where the first radio observation was obtained 1.4 days after explosion
%and 2019ein, is needed to constrain a V3890 Sgr-like progenitor 
(\citealt{Horesh+12}, see also \citealt{Chomiuk2012}). 
%This is due to the fact that these studies are primarily based on observations of individually classified SNe Ia even though the progenitors may be  quite diverse. 
A symbiotic progenitor with a wind mass loss rate $> 2 \times 10^{-8} ~\textrm M_\odot~\textrm{yr}^{-1}$ is ruled out for SN\,2011fe, assuming $\epsilon_e = 0.1$, $\epsilon_B = 0.01$, and $V_{\textrm {wind}} = 10$ km s$^{-1}$ \citep{Horesh+12}. 
This radio limit from $1.4$ days after explosion constrains the CSM at a radius $\sim 1.5 \times 10^{15}$ cm. Given that the radio light curve of V3890~Sgr is well modeled as an $r^{-2}$ stellar wind with $\dot{M} = 3.2 \times 10^{-8} ~\textrm M_\odot~\textrm{yr}^{-1}$ out to day $30$ (shock radius $\sim 8 \times 10^{14}$ cm; Figure \ref{fig:Radius_Velocity_evolution}) and then a shell of excess CSM density out to day $\sim$100 ($\sim 2 \times 10^{15}$ cm), %and taking the same microphysical parameter values as used in SN\,2011fe (i.e., model E in Table \ref{tab:aconf}),
we can rule out a V3890~Sgr progenitor to SN\,2011fe. In the future, such direct comparisons with real-world candidate progenitors can be extended to additional SNe Ia by modeling the blastwave evolution in embedded novae like V3890~Sgr (constrained by optical spectral line profiles) and by scaling-up the radio light curves observed for novae to the blastwave energetics expected for SNe Ia.

%is needed to compare the full radio light curve of V3890 Sgr with the later radio limits obtained for many SNe Ia

%we can see that 
%These same authors state that the radio upper limits may not be sufficient to rule out novae that undergo recurrent thermonuclear eruptions \citep{Pellegrino2020}. 
%Currently favored progenitor scenarios of SN Ia with circumstellar material created by novae shells typically have mass-loss rates in the range of $(10^{-9} \leq \dot M \leq 10^{-7})~\textrm M_\odot~\textrm{yr}^{-1}$ for $V_\textrm{wind} > 10~\textrm{km~s}^{-1}$ %depending on the velocity of the wind near the white dwarf
%\citep{Chomiuk2012, Lundqvist20}. %This is consistent with the mass loss rate estimated for V3890 Sgr estimated in \S \ref{subsec:modelling}.
%The red giant mass-loss rates for this system is ruled out as a Ia progenitor

%A small number of SNe have shown the presence of H$\alpha$ emission lines and Na{\sc~i} D absorption lines, which have been interpreted as the presence of circumstellar material around them \citep[e.g.,][]{patat2007}. A similar evolution of variable Na{\sc~i} D absorption features for RS Oph is presented by \citet{patat2011}. The circumstellar material around SN Ia PTF 11kx is also shown to be structured such that regions of less dense material alternate with those of higher density material, commonly referred to as cavities and shells, respectively \citep{Dilday2012}. Based on the complex surrounding medium of recurrent symbiotic novae \citep{Walder2008, Booth2016}, it is possible that some of these systems are potential progenitors of SNe Ia. 

\section{Conclusions}
\label{sec:conclusions_V3890Sgr}

The symbiotic recurrent nova V3890 Sgr was observed for $18$ months at radio frequencies with MeerKAT and the VLA (Figure~\ref{fig:opticallightcurve}). The radio emission is detected from $3$ days after optical discovery and peaks between days $11$ and $15$ at all observed frequencies. The rapid time evolution of the radio light curve, steep spectral indices (Figure~\ref{fig:radio_light_curve}) and high brightness temperatures (Figure~\ref{fig:V3890Sgr_T_B}) are indicators of synchrotron radiation from the nova, similar to what is observed in other symbiotic recurrent novae \citep[e.g.,][]{Kantharia2016}.

The synchrotron emission in V3890 Sgr is suggested to arise from the interaction between the ejected nova envelope and a structured circumbinary medium formed by the pre-existing wind from the red giant star. The radio light curve is characterised by a secondary peak following the initial peak, 
%at $\tau = 1$, 
and then decays faster than expected for a shock wave propagating in a windlike medium ($\rho_{\textrm CSM} \propto r^{-2}$; Figure~\ref{fig:modellightcurve1}). This hints at a complex structure of circumstellar medium around the nova, with an excess of material at $\sim 10^{15}$ cm surrounded by a relatively low density outer environment. 
%This could be similar to other novae where the eruption occurs inside a red giant wind such as the recurrent nova RS Oph.

The turn-on of the radio emission fits well with a model of decreasing opacity of the ionized circumstellar gas as the ejected envelope of $\approx10^{-6}~\textrm{M}_\odot$ %shock wave
sweeps through it \citep{chevalier1981}. Modelling of radio emission provides a mass-loss rate of the red giant companion star on the order of ${\dot M}=3.2\times 10^{-8}~\textrm M_\odot~\textrm{yr}^{-1}$ for $V_\textrm{wind} = 10~\textrm{km~s}^{-1}$ and a distance of $9$ kpc. Similar mass-loss rates are estimated for RS Oph \citep{vanLoon2008} and other symbiotic stars \citep{Seaquist1990}. By comparing our light curve with radio upper limits for SNe Ia, we conclude that a V3890~Sgr-like progenitor can be ruled out for the well-observed SN\,2011fe. 

%The estimated mass-loss rate and possible structural distribution of the circumbinary material as determined from the radio light curve presented here, are consistent with the CSM parameter space of SNe Ia allowed by radio limits.

%The last numbered section should briefly summarise what has been done, and describe
%the final conclusions which the authors draw from their work.

\section*{Acknowledgements}

%The Acknowledgements section is not numbered. Here you can thank helpful
%colleagues, acknowledge funding agencies, telescopes and facilities used etc.
%Try to keep it short.
%We thank the referee for providing useful comments?..........
The MeerKAT telescope is operated by the South African Radio Astronomy Observatory (SARAO; \url{www.ska.ac.za}), which is a facility of the National Research Foundation (NRF), an agency of the Department of Science and Innovation. The data was processed using the Ilifu cloud computing facility –
\url{www.ilifu.ac.za}, a partnership between the University of Cape Town, the University of the Western Cape, the University of Stellenbosch, Sol Plaatje University, the Cape Peninsula University of Technology, and the South African Radio Astronomy Observatory. The Ilifu facility is supported by contributions from the Inter-University Institute for Data Intensive Astronomy (IDIA – \url{https://www.idia.ac.za/} a partnership between the University of Cape Town, the University of Pretoria and the University of the Western Cape), the Computational Biology division at UCT, and the Data Intensive Research Initiative of South Africa (DIRISA)). This research was supported by SARAO student bursary. MMN and PAW kindly acknowledge financial support from the University of Cape Town and the NRF Grant $129359$ and NRF SARChI Grant $111692$. 
LC is grateful for NSF support through grants AST-1751874, AST-1907790, and AST-2107070.
VARMR acknowledges financial support from the Funda\c{c}\~{a}o para a Ci\^encia e a Tecnologia (FCT) in the form of an exploratory project of reference IF/00498/2015/CP1302/CT0001, and supported by Enabling Green E-science for the Square Kilometre Array Research Infrastructure (ENGAGE-SKA), POCI-01-0145-FEDER-022217, funded by Programa Operacional Competitividade e Internacionaliza\c{c}\~ao o (COMPETE 2020) and FCT, Portugal.
%Obligatory NRAO acknowledgement, from Justin
We thank the NRAO for the generous allocation of time on on the VLA. The National Radio Astronomy Observatory is a facility of the National Science Foundation operated under cooperative agreement by Associated Universities, Inc.

%%%%%%%%%%%%%%%%%%%%%%%%%%%%%%%%%%%%%%%%%%%%%%%%%%
\section*{Data Availability}

 %The inclusion of a Data Availability Statement is a requirement for articles published in MNRAS. Data Availability Statements provide a standardised format for readers to understand the availability of data underlying the research results described in the article. The statement may refer to original data generated in the course of the study or to third-party data analysed in the article. The statement should describe and provide means of access, where possible, by linking to the data or providing the required accession numbers for the relevant databases or DOIs.
All MeerKAT data for V3890 Sgr are available from the MeerKAT archive at \href{https://archive.sarao.ac.za/}{archive.sarao.ac.za} search under the project codes SCI-20190418-MN-01 and DDT-20200323-MN-01.
%--Justin--
All VLA data for this project are publicly available from the NRAO archive at \href{https://data.nrao.edu}{data.nrao.edu}. The project codes were 19A-313 and 20B-302.

%%%%%%%%%%%%%%%%%%%% REFERENCES %%%%%%%%%%%%%%%%%%

% The best way to enter references is to use BibTeX:

\bibliographystyle{mnras}
\bibliography{References} % if your bibtex file is called example.bib

% Alternatively you could enter them by hand, like this:
% This method is tedious and prone to error if you have lots of references
%\begin{thebibliography}{99}
%\bibitem[\protect\citeauthoryear{Author}{2012}]{Author2012}
%Author A.~N., 2013, Journal of Improbable Astronomy, 1, 1
%\bibitem[\protect\citeauthoryear{Others}{2013}]{Others2013}
%Others S., 2012, Journal of Interesting Stuff, 17, 198
%\end{thebibliography}

%%%%%%%%%%%%%%%%%%%%%%%%%%%%%%%%%%%%%%%%%%%%%%%%%%

%%%%%%%%%%%%%%%%% APPENDICES %%%%%%%%%%%%%%%%%%%%%

%\appendix

%\section{Some extra material}

%If you want to present additional material which would interrupt the flow of the main paper,
%it can be placed in an Appendix which appears after the list of references.

%%%%%%%%%%%%%%%%%%%%%%%%%%%%%%%%%%%%%%%%%%%%%%%%%%

% Don't change these lines
\bsp	% typesetting comment
\label{lastpage}
\end{document}